\documentclass[journal]{IEEEtran}
\usepackage{graphicx}
\usepackage{array}
\usepackage{epsfig}
\usepackage{epstopdf}
\usepackage{lipsum,multicol}
\usepackage{cite}
\usepackage[ruled,vlined,linesnumbered,resetcount]{algorithm2e}

\usepackage{mathtools}
\usepackage{breqn}
\usepackage{cuted}
\usepackage{blindtext}
\usepackage{wrapfig}
\usepackage{url}
\usepackage{stfloats}

\usepackage{soul}
\usepackage{xcolor}

\usepackage{longtable}
\usepackage{booktabs}

\usepackage{caption}
\usepackage{subcaption}

\usepackage{enumitem}
\usepackage{amsmath,amssymb,amsfonts,amsthm,bm}
\usepackage{multirow}

\usepackage{graphicx}
\usepackage{ifluatex}
\ifluatex
  \usepackage{ifpdf}
  \ifpdf
    \expandafter\ifx\csname Gin@rule@.jp2\endcsname\relax
      \usepackage{grfext}
      \AppendGraphicsExtensions{.jp2,.JP2}
    \fi
  \fi  
\fi

\makeatother

\usepackage{caption,subcaption}

\ifCLASSINFOpdf
\else
\fi
\begin{document}
\bstctlcite{IEEEexample:BSTcontrol}

This paper was submitted for publication on the \textit{IEEE Transactions on Communications} on May 12, 2022  and was assigned reference number TCOM-TPS-22-0564.
It was finally accepted for publication on November 26, 2022.  

\bigskip

\copyright 2022 IEEE. Personal use of this material is permitted. Permission from IEEE must be obtained for all other uses, in any current or future media, including reprinting/republishing this material for advertising or promotional purposes, creating new collective works, for resale or redistribution to servers or lists, or reuse of any copyrighted  component of this work in other works.

%
\title{ \Large Power Control in Cell-Free Massive MIMO Networks for UAVs URLLC under the Finite Blocklength Regime}
%
%
%

 \author{Mohamed~Elwekeil,~\IEEEmembership{}
         Alessio~Zappone,~\IEEEmembership{Senior Member,~IEEE,}
         and~Stefano~Buzzi,~\IEEEmembership{Senior Member,~IEEE} \vspace{-0.75cm}
 \thanks{M. Elwekeil, A. Zappone, and S. Buzzi are with the Department of Electrical and Information Engineering, University of Cassino and Southern Lazio, Cassino, Italy. (email:\{mohamed.elwekeil, alessio.zappone, s.buzzi\}@unicas.it).}
 \thanks{M. Elwekeil is on leave from Department of Electronics and Electrical Communications Engineering, Faculty of Electronic Engineering, Menoufia University, Menouf 32952, Egypt. (email: mohamed.elwekeil@el-eng.menofia.edu.eg).}
\thanks{A. Zappone, and S. Buzzi are also with Consorzio Nazionale Interuniversitario per le Telecomunicazioni, Parma, Italy.} 
\thanks{S. Buzzi is also with Politecnico di Milano, Milano, Italy.}
\thanks{This work was supported by H2020 Marie Skłodowska-Curie Actions (MSCA) Individual Fellowships (IF) IUCCF, grant 844253. The work of S. Buzzi was also partially supported by the MIUR PRIN 2017 Project "LiquidEdge" and by the MIUR Project "Dipartimenti di Eccellenza 2018-2022.}  
 }

\maketitle

 \begin{abstract}
In this paper, we employ a user-centric (UC) cell-free massive MIMO (CFmMIMO) network for providing ultra reliable low latency communication (URLLC) when traditional ground users (GUs) coexist with unmanned aerial vehicles (UAVs). We study power control in both the downlink and the uplink when partial zero-forcing (PZF) transmit/receive beamforming and maximum ratio transmission/combining are utilized. We consider optimization problems where the objective is to maximize either the users' sum URLLC rate or the minimum user's URLLC rate. The URLLC rate function is both complicated and nonconvex rendering the considered optimization problems nonconvex. Thus, we propose two approximations for the complicated URLLC rate function and employ successive convex optimization (SCO) to tackle the considered optimization problems. Specifically, we propose the SCO with iterative concave lower bound approximation (SCO-ICBA) and the SCO with iterative interference approximation (SCO-IIA). We provide extensive simulations to evaluate SCO-ICBA and SCO-IIA and compare UC CFmMIMO deployment with traditional colocated massive MIMO (COmMIMO) systems. The obtained results reveal that employing the SCO-IIA scheme to optimize the minimum user's rate for CFmMIMO with MRT in the downlink, and PZF reception in the uplink can provide the best URLLC rate performances. 
 \end{abstract}



%
\IEEEpeerreviewmaketitle

\section{Introduction}
%
 \IEEEPARstart{R}{ecently,} the wireless communications research has witnessed a huge development to accommodate the continuous increase in the data traffic demand and the emergence of new use cases. Specifically, in the era of the internet of things (IoT), the number of connected devices is increasing and new services are evolving. IoT has applications in many fields such as health care, manufacturing, transportation, entertainment, etc \cite{chen2014vision}. Wireless communication is one of the main enablers for a successful IoT system \cite{ijaz2016enabling}. Thus, the forthcoming wireless networks will not only increase the data rates for mobile broadband (MBB) applications, but also, satisfy the requirements of the emerging IoT applications \cite{brito2016trends,schulz2017latency}. 

Consequently, 5G and beyond-5G networks will have to deal with a very wide range of requirements which include massive connectivity, huge data rate, ultra high reliability, and very small round trip latency (in the range of 1ms) \cite{hassan2021key}. Particularly, ultra reliable low latency communication (URLLC) will be crucial for various mission-critical applications such as autonomous driving, remote surgery, tactile internet, catastrophe rescue, haptic feedback, augmented/virtual reality, and unmanned aerial vehicle (UAV) communications \cite{bennis2018ultrareliable}. UAVs were originally utilized for military applications. Recently, they have been used for civil applications. It has been proposed to utilize the UAVs in wireless communications \cite{zeng2016wireless}. Utilizing UAVs along with colocated massive MIMO (COmMIMO) has been studied in the literature. For example, COmMIMO is used as an infrastructure to provide connectivity for UAVs and traditional ground users (GUs) \cite{geraci2018supporting, huang2021massive}.  

The finite blocklength communication (FBLC) regime attracts the attention of the wireless communications research community to fulfill the requirements of URLLC applications \cite{polyanskiy2010channel, yang2014quasi, mary2016finite}. Most of the works in the literature consider cellular networks when studying URLLC. For instance, the work in \cite{nasir2020resource} has investigated the resource allocation problem in the FBLC regime for cellular networks considering both single antenna base station (BS), and multiple antennas BS. However, the authors in \cite{nasir2020resource} studied only the case of single cell. The authors in \cite{hu2018optimal,hu2018optimal2} have considered the power allocation for a downlink broadcasting scenario where the FBLC regime is employed in a time division multiple access (TDMA) system. However, the works in both \cite{hu2018optimal,hu2018optimal2} concentrated on the single cell scenario where a single BS is transmitting URLLC packets to multiple users.

The work in \cite{ren2020joint} studied joint pilot and payload uplink power allocation problem for URLLC in an industrial IoT network supported by COmMIMO. However, the authors of \cite{ren2020joint} considered only a single cell COmMIMO system where multiple users are transmitting their URLLC packets to a single COmMIMO BS. The optimization of URLLC transmission by minimizing the latency, while satisfying reliability constraints and power limits, in an autonomous driving scenario where a sensor node monitoring the traffic so as to warn the neighboring vehicles through the road side units about any emergency is considered in \cite{shen2018joint}. The authors in \cite{she2019ultra} considered URLLC for a scenario where UAVs are controlled by a ground station that is supported by a number of APs. Specifically, they optimized the altitude of the UAVs, the uplink transmission duration, the downlink transmission duration, the number of APs, and the number of antennas in each AP so as to maximize the converge range of the ground station. However, the authors in \cite{she2019ultra} assumed that orthogonal resources are allocated to different UAVs and thus they did not consider the impact of interference. Besides, they did not consider the power allocation problem for the UAVs; instead, they assumed uniform power allocation. Also, they employed the APs only to serve the UAVs and thus did not consider the existence of any GUs. The employment of an UAV as a relay to enable URLLC in an industrial automation scenario has been studied in \cite{ren2020joint2}, where the location of the UAV-relay and transmission power have been optimized to maximize the signal to noise ratio at the receiver side.

Cell-free massive MIMO (CFmMIMO) is a promising technology that provides sufficiently good service to almost every user in the network \cite{ngo2015cell}. The basic idea of CFmMIMO is to distribute the antennas in the service area by deploying many small access points (APs), instead of collecting the antennas at a BS in COmMIMO. The authors of \cite{ngo2017cell} compared CFmMIMO and small cells, and showed that CFmMIMO provides better throughput performance. In \cite{buzzi2017cell}, a user-centric (UC) approach for CFmMIMO was presented where the user should be served by only a subset of the APs based on APs' distances to the user. The channel hardening of CFmMIMO system has been investigated in \cite{polegre2020new, polegre2020channel}. More details about the analysis and design of CFmMIMO can be found in \cite{polegre2021analysis}. The work in \cite{masoumi2021coexistence} studied the coexistence of device-to-device (D2D) communication and CFmMIMO in the uplink. The authors of \cite{ngo2017total} considered the energy efficiency of CFmMIMO, where they presented user-AP association and power control algorithms so as to enhance the energy efficiency of the CFmMIMO network. The authors in \cite{d2019cell, 8952782} investigated the employment of CFmMIMO for supporting UAV communications. The obtained results in \cite{d2019cell, 8952782} declared that CFmMIMO can provide better performance than COmMIMO. However, the authors in \cite{d2019cell, 8952782} did not consider the FBLC regime. In \cite{elwekeil2021optimal}, a joint beamforming and power control scheme in the downlink of CFmMIMO is proposed. Most of the works on CFmMIMO assume broadband applications where reliability and low latency are not the targets; thus they did not consider the FBLC regime (e.g.,\cite{ngo2015cell, ngo2017cell, buzzi2017cell, polegre2020new, polegre2020channel, polegre2021analysis, masoumi2021coexistence, ngo2017total, d2019cell, 8952782, elwekeil2021optimal}).

FBLC regime in CFmMIMO is rarely studied in the literature. For example, \cite{nasir2021cell} presented power control scheme for the CFmMIMO system that employs FBLC for providing URLLC. The authors in \cite{nasir2021cell} adopted a special class of the conjugate beamforming (CB) and found the power control coefficients in a CFmMIMO system. However, the authors employed a single antenna at every AP in the CFmMIMO system and considered the power control for the downlink only. Furthermore, they did not consider the existence of the UAVs. Moreover, they assumed perfect channel knowledge which is not practical in real communication systems. In their evaluations, they only considered the CFmMIMO and did not benchmark it with COmMIMO. The performance of CFmMIMO for providing URLLC services is investigated in \cite{lancho2021cell}, where the authors studied the impact of the number of antennas employed by the system on the network availability. However, they did not provide any power allocation schemes. In \cite{peng2022resource}, the authors presented a power control scheme for only the downlink of CFmMIMO system. However, the authors of \cite{peng2022resource} considered a simple system that has few users (5 - 15 users) and few APs (1, 4, or 9 APs). We thereby emphasize that the research in the topic of FBLC regime for URLLC over CFmMIMO is still in its infancy and has not reached the maturity stage; thus, more investigations in this topic are needed. 
 
In this work, we deal with the power control in both the uplink and the downlink of a UC CFmMIMO network that provides URLLC for both traditional GUs and UAVs by employing the FBLC regime. For the sake of brevity we refer to the UC CFmMIMO by only CFmMIMO in the rest of the paper. We formulate optimization problems in order to either maximize the users' sum URLLC rate or maximize the minimum user's URLLC rate. The resulting optimization problems are nonconvex \cite{boyd2004convex}. 
Thus, we present two schemes to approximate the user's URLLC rate function and employ an iterative method to tackle the considered optimization problems. We study a special class of zero-forcing (ZF) transmission/reception at the APs, namely, the partial zero-forcing (PZF) \cite{interdonato2020local}. Specifically, the existence of multiple antennas at each receiving/transmitting AP will be employed to remove only the set of the strongest interference components. Besides, we consider the maximum ratio transmission (MRT) and maximum ratio combining (MRC) at APs, which are commonly employed in the literature for both downlink and uplink, respectively. We use the proposed schemes to approximate the sophisticated FBLC rate expression into two distinct concave forms; thus, the considered optimization problems can be approximated to convex forms where we employ successive convex optimization (SCO) \cite{razaviyayn2014successive,scutari2013decomposition} to find a local solution for the uplink/downlink power control coefficients. To the best of our knowledge, this is the first work to employ CFmMIMO network to support the coexistence of UAVs and traditional GUs under the FBLC regime for URLLC. 

The contributions of this work are summarized as follows:

\begin{itemize}

\item

The paper proposes two approximations to tackle the power control problem in CFmMIMO networks presenting URLLC services for both UAVs and traditional GUs. The proposed approximations are not restricted to the CFmMIMO as it can be employed for different network architectures including COmMIMO. The considered scenario is quite challenging, where we can not control the UAVs' locations because the UAVs are assumed to be hovering to perform a certain mission.

\item 

The power control problem is formulated in two distinct optimization problems with objectives so as to either maximize the URLLC users' sum rate, or maximize the worst URLLC user's rate. Since the URLLC  rate expression is a sophisticated nonconcave function of the power control coefficients, the considered optimization problems are intractable using traditional optimization tools. Thus, we solve these optimization problems iteratively using SCO. Therefore, we can obtain a sequence of progressively improving feasible solutions that can converge to a locally optimal solution.  

\item

We perform extensive simulations to evaluate the proposed two schemes for both uplink and downlink URLLC considering different network architectures. Specifically, we compare the performance of both CFmMIMO and COmMIMO when both are employed to provide URLLC under the FBLC regime for UAVs coexisting with GUs. In our comparison, we benchmarked the ZF scheme with the MRC in the uplink and MRT in the downlink.

\end{itemize}

%
%

 

\section{System Model} \label{sec:system}
\subsection{Cell-Free Massive MIMO Network}
\begin{figure}[!t]
\begin{center}
\captionsetup{font=small}
\includegraphics[width =0.7\columnwidth] {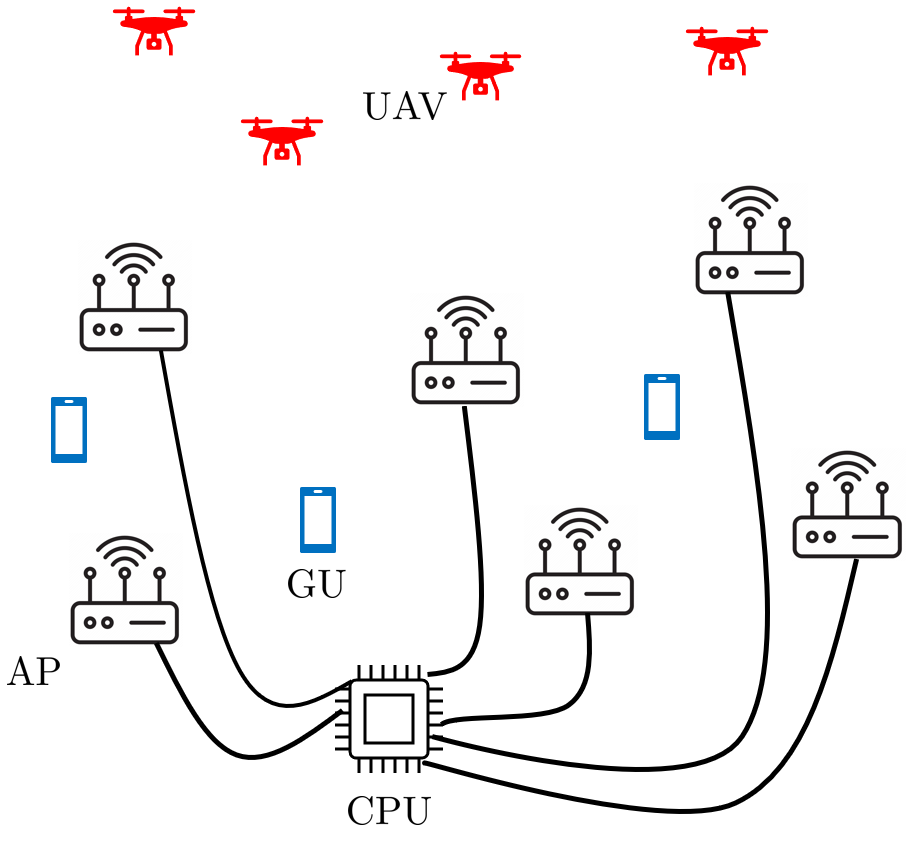}
\caption{The considered system model.}\vspace{-0.4cm}
\label{fig:system model}
\end{center}
\end{figure}
The CFmMIMO system model is shown in Fig. \ref{fig:system model}. Specifically, we assume that the FBLC regime is employed for providing URLLC in a CFmMIMO network that concurrently serves both UAVs hovering on the horizon for a specific mission and traditional GUs. We treat the UAVs as flying clients where we have no control on their locations that have been optimized beforehand for a certain mission, rather we concentrate on employing the CFmMIMO to serve these UAVs along with the traditional GUs. 
Assume that the CFmMIMO network incorporate a set of access points $\mathbb{A}=\{AP_1,AP_2,..,AP_a,..,AP_{N_A}\}$, where $N_A=\#\{\mathbb{A}\}$ represents the cardinality of the set $\mathbb{A}$, i.e., the number of APs in the CFmMIMO network. Here, $a$ is the index of the $a^{th}$ AP. We consider that all APs are connected to a central processing unit (CPU) through ideal fronthaul connections. Let $\mathbb{V}=\{UAV_1,UAV_2,..,UAV_v,..,UAV_{N_V}\}$ denotes the set of the UAVs served by the CFmMIMO network, where $N_V=\#\{\mathbb{V}\}$ is the cardinality of the set $\mathbb{V}$, which is the number of the served UAVs, and $v$ is the index of the $v^{th}$ UAV. Besides, let $\mathbb{G}=\{GU_1,GU_2,..,GU_g..,GU_{N_G}\}$ stands for the set of the ground users served by the CFmMIMO, where $N_G=\#\{\mathbb{G}\}$ is the cardinality of the set $\mathbb{G}$, i.e., the number of the served ground users, and $g$ is the index of the $g^{th}$ ground user. We hereby represent by $\mathbb{U}$ the set of all users, including both UAVs and GUs, that are served by the CFmMIMO network, i.e., $\mathbb{U}=\mathbb{G} \cup \mathbb{V}=\{U_1,U_2,..,U_u..,U_{N_U}\}$, where $u$ is the index of the $u^{th}$ user. Here, the total number of all served users in the network is $N_U=\#\{\mathbb{U}\} = \#\{\mathbb{G}\}+\#\{\mathbb{V}\}=N_G+N_V$. 

Every AP in the CFmMIMO network is equipped with $M_{AP}$ antennas, while every served user, either an UAV or a GU, employs a single antenna. In this work, we follow the user-AP association scheme frequently adopted in the literature (e.g., \cite{ngo2017total,8952782}). Particularly, every user $u$ is served by a subset of the whole available APs according to the large scale fading coefficients. The service is provided to a certain user $u$ by a subset of the existing APs, $\mathbb{A}_{u}=\{AP_1,AP_2,..,AP_{N_{A,u}}\}$, where $N_{A,u}$ represents the number of APs that provide the service to the user $u$, and the set $\mathbb{A}_{u}$ contains the $N_{A,u}$ APs that have the highest large scale fading coefficients to this user. Consequently, we use $\mathbb{U}_a=\{U_1, U_2,...,U_{N_{U,a}}\}$ to represent the set of users served by AP $a$. 

\subsection{TDD Transmission}
 In alignment with the convention employed in the CFmMIMO literature (e.g., \cite{ngo2015cell, ngo2017cell, buzzi2017cell, ngo2017total, d2019cell, 8952782}), we adopt the time division duplex (TDD) transmission framework, where the whole communication process occurs on the same frequency band. Assume that the length of the coherence interval is $\tau_c$ time-frequency samples.  This coherence interval will be divided into three sub-intervals. The first sub-interval of length $\tau_p<\tau_c$ is employed for uplink channel estimation, where every user $u \in \mathbb{U}$ transmits a pilot sequence so that every AP in the network can perform channel estimation. The second sub-interval of length $\tau_d<\tau_c$ is used for the downlink data transmission, where every AP $a$ transmits the downlink data to the users it serves, i.e., $u \in\mathbb{U}_a$. Finally, the third sub-interval of length $\tau_u<\tau_c$ is utilized for the uplink  data transmission, where every user transmits its uplink data. The lengths of the aforementioned sub-intervals should be selected such that $\tau_p+\tau_d+\tau_u \leq \tau_c$. In this paper, we assume that equal length sub-intervals are employed for uplink data transmission and downlink data transmission, i.e., $\tau_d=\tau_u=\displaystyle \frac{\tau_c-\tau_p}{2}$ . 
During the channel estimation phase, we employ long pilot sequences to mitigate the impact of pilot contamination. Thus, we can reliably estimate the channel during the coherence interval of interest. Note that the coherence interval $\tau_c$ and consequently the sub-intervals $\tau_u$ and $\tau_d$ are quite longer than the duration of each corresponding URLLC transmission. Thus, multiple uplink and downlink URLLC transmissions are executed during the intervals $\tau_u$ and $\tau_d$, respectively.

\subsection{Channel Model}
The channel vector from the $u^{th}$ user to the $a^{th}$ AP in the coherence interval of interest is denoted by $\boldsymbol{h}_{u,a}\in \mathbb{C}^{M_{AP}\times 1}$. For the UAVs, we consider that the small scale fading is Rician distributed, where the channel is composed of multiple paths with only one line of sight (LOS) path. On the other hand, for the GUs, we consider that the small scale fading is Rayleigh distributed, which can be considered as a special case of the Rician fading with Rician K-factor that equals zero. Thus, $\boldsymbol{h}_{u,a}$ can be expressed as in \cite{8952782}
\begin{equation}
\label{eq:eq-1}
 \begin{array}{ll}
\boldsymbol{h}_{u,a}& = \sqrt{\frac{\beta_{u,a}}{K_{u,a}+1}}\left[\sqrt{K_{u,a}} e^{j\mu_{u,a}}\boldsymbol{a}(\theta_{u,a})+\boldsymbol{g}_{u,a} \right],
\end{array}
\end{equation}
where $\beta_{u,a}\in \mathbb{R}$ is the large scale fading coefficient, which includes the impacts of the path loss and the shadowing, between user $u$ and AP $a$. Besides, $K_{u,a}$ is the Rician K-factor between user $u$ and AP $a$, $\mu_{u,a}$ is the phase offset of the direct path between user $u$ and AP $a$, $\theta_{u,a}$ is the reference angle of the direct path from user $u$ to AP $a$, and $\boldsymbol{a}(\theta_{u,a})\in \mathbb{C}^{M_{AP}\times 1}$ is the steering vector at angle $\theta_{u,a}$ \cite{8952782}. In addition, $\boldsymbol{g}_{u,a}\in \mathbb{C}^{M_{AP}\times 1} $ is the vector of independent and identically distributed (i.i.d.) small scale fading coefficients between user $u$ and AP $a$; i.e., $\boldsymbol{g}_{u,a} \sim \mathcal{CN}(0,\boldsymbol{I}_{M_{AP}})$, where $\boldsymbol{I}_{M_{AP}}$ is the $M_{AP} \times M_{AP}$ identity matrix.
 
 \subsection{Channel Estimation}
We assume that the linear minimum mean square error (LMMSE) channel estimation is employed at each AP similar to \cite{ngo2017cell,8952782}. Assume that $\bm{\psi}_u \in \mathbb{C}^{\tau_{p}\times 1}$ is the pilot sequence transmitted by user $u$, where $\left \|\bm{\psi}_u  \right \| ^2=1$. Let $\alpha_{p,u}$ denotes the uplink pilot transmit power sent by user $u$ in the training phase, then the transmitted signal from user $u$ during the training phase can be represnted by $\sqrt{\alpha_{p,u}}\bm{\psi}_u^*$. Thus, the corresponding component arriving at the AP $a$ is $\sqrt{\alpha_{p,u}}\boldsymbol{h}_{u,a}\bm{\psi}_u^*$. $\mathbf{\Phi}_a\in \mathbb{C}^{M_{AP}\times \tau_{p}}$ denotes the cumulative received pilot signal at AP $a$ during the uplink training phase, which is given by
\begin{equation}
\label{eq:eq-1_training}
\begin{array}{ll}
\mathbf{\Phi}_a & = \displaystyle \sum_{u \in\mathbb{U}} \sqrt{\alpha_{p,u}}\boldsymbol{h}_{u,a}\bm{\psi}_u^* +\boldsymbol{W}_{a,p},
\end{array}
\end{equation} 
where $\boldsymbol{W}_{a,p}\in \mathbb{C}^{M_{AP}\times \tau_{p}}$ accounts for both the AWGN and external interference at AP $a$ during the training phase. The components of $\boldsymbol{W}_{a,p}$ are i.i.d. complex Gaussian random variables, i.e., each component follows $\mathcal{CN}(0,\sigma_w)$. Here, the operator $(.)^*$ represents the conjugate transpose operation.

Then, the AP $a$ can multiply $\mathbf{\Phi}_a $ by the pilot sequence utilized by the user $u$ to calculate the following statistics \cite{8952782}
\begin{equation}
\label{eq:eq-2_training}
\begin{array}{ll}
\bm{\hat{\phi}}_{u,a} & = \mathbf{\Phi}_a\bm{\psi}_u 
\\ & = \sqrt{\alpha_{p,u}}\boldsymbol{h}_{u,a}+\boldsymbol{W}_{a,p}\bm{\psi}_u + \displaystyle \sum_{i \in\mathbb{U}\setminus u} \sqrt{\alpha_{p,i}}\boldsymbol{h}_{i,a}\bm{\psi}_i^* \bm{\psi}_u.
\end{array}
\end{equation}
The LMMSE channel estimation for the channel from user $u$ to the AP $a$ in the coherence interval of interest, $\boldsymbol{\hat{h}}_{u,a}$, can be calculated accordingly from the statistics $\bm{\hat{\phi}}_{u,a}$ as follows \cite{8952782}
\begin{equation}
\label{eq:eq-3_training}
\begin{array}{ll}
\boldsymbol{\hat{h}}_{u,a} = \boldsymbol{D}_{u,a}\bm{\hat{\phi}}_{u,a},
\end{array}
\end{equation}
where $\boldsymbol{D}_{u,a} \in \mathbb{C}^{M_{AP}\times M_{AP}}$ is given by
\begin{equation}
\label{eq:eq-4_training}
\begin{array}{ll}
\boldsymbol{D}_{u,a} = \sqrt{\alpha_{p,u}}\boldsymbol{C}_{u,a}\boldsymbol{B}_{u,a}^{-1},
\end{array}
\end{equation}
with $\boldsymbol{C}_{u,a}$ given by
\begin{equation}
\label{eq:eq-5_training}
\begin{array}{ll}
\boldsymbol{C}_{u,a} = \frac{\beta_{u,a}}{K_{u,a}+1}\left[ K_{u,a}\boldsymbol{a}(\theta_{u,a})\boldsymbol{a}^*(\theta_{u,a}) + \boldsymbol{I}_{M_{AP}}\right]
\end{array}
\end{equation}
and $\boldsymbol{B}_{u,a}$ is given by
\begin{equation}
\label{eq:eq-6_training}
\begin{array}{ll}
\boldsymbol{B}_{u,a} = \displaystyle \sum_{i \in\mathbb{U}}\alpha_{p,u}\boldsymbol{C}_{i,a} \left |\bm{\psi}_i^* \bm{\psi}_u \right |^2 + \sigma_w^2\boldsymbol{I}_{M_{AP}}
\end{array}
\end{equation}

It should be noted that the utilized channel estimation is dependent on $\bm{\hat{\phi}}_{u,a}$, which can differ slightly in every coherence interval. In addition, it is dependent on $\boldsymbol{D}_{u,a}$ which relies on the large scale fading parameters $\beta_{u,a}$ and $K_{u,a}$, along with $\boldsymbol{a}(\theta_{u,a})$ that change together on a relatively long time scale. The data detection process, which is not the main focus of this paper, is very sensitive to instantaneous channel state information (CSI) that is synchronized with $\bm{\hat{\phi}}_{u,a}$. However, the proposed power control scheme does not require the instantaneous CSI, where the power control can be only updated on the same time scale at which the large scale fading changes, which can include multiple coherence intervals. Let's assume that the large scale fading parameters changes every $L$ coherence intervals, and the channel estimation in the first coherence interval is denoted by $\boldsymbol{\hat{h}}_{LS;u,a} = \boldsymbol{D}_{u,a}\bm{\hat{\phi}}_{LS;u,a}$. Therefore, we can utilize this $\boldsymbol{\hat{h}}_{LS;u,a}$ for the power control of all URLLC transmissions in both the uplink and downlink of the successive $L$ coherence intervals that have the same $\boldsymbol{D}_{u,a}$, i.e., large scale fading parameters. In other words, the proposed power control schemes do not need exchanging the estimated channels every coherence interval. 
  
\subsection{Downlink Data Transmission}
In this part, we concentrate on the downlink where the APs in the CFmMIMO system are transmitting to multiple users including UAVs and traditional GUs. The $n^{th}$ signal transmitted from a certain AP $a$, $\boldsymbol{x}_a$ is given by
\begin{equation}
\label{eq:eq-1_DL}
\begin{array}{ll}
\boldsymbol{x}_{a}& = \displaystyle \sum_{u \in\mathbb{U}_a}\sqrt{\alpha_{a,u}^{dl}}\boldsymbol{f}_{a,u}s_u^{dl},
\end{array}
\end{equation}
where $\boldsymbol{f}_{a,u}\in \mathbb{C}^{M_{AP}\times 1}$ is the corresponding downlink beamforming vector, and $s_u^{dl}$ is the $n^{th}$ downlink data symbol for user $u$. Note that, we employ the normalized beamforming scheme, i.e., $ \left \| \boldsymbol{f}_{a,u} \right \|^2=1$, where  $\boldsymbol{f}_{a,u}$ determines only the spatial direction and $\alpha_{a,u}^{dl}$ is the corresponding downlink power control coefficient from AP $a$ to user $u$.

The received signal at user $u$ is denoted by $y_u$, which is expressed as 
\begin{equation}
\label{eq:eq-3_DL}
\begin{array}{ll}
y_u& = \displaystyle \sum_{a \in\mathbb{A}}\boldsymbol{h}_{a,u}\boldsymbol{x}_{a}+w_u
\\
  & = \displaystyle \sum_{a \in\mathbb{A}} \displaystyle \sum_{i \in\mathbb{U}_a}\sqrt{\alpha_{a,i}^{dl}}\boldsymbol{h}_{u,a}^*\boldsymbol{f}_{a,i}s_i^{dl}+w_u,
\end{array}
\end{equation} 
where $w_u\sim \mathcal{CN}(0,\sigma_u^2)$ corresponds to the AWGN at user $u$, and $\sigma_u^2$ represents the noise variance. We hereby reformulate equation (\ref{eq:eq-3_DL}) by switching the summations as in equation (\ref{eq:eq-4_DL}) where the desired signal term can be separated from the interference terms at user $u$.
\begin{equation}
\label{eq:eq-4_DL}
\begin{array}{ll}
y_u&= \underbrace{\displaystyle \sum_{a \in\mathbb{A}_u}\sqrt{\alpha_{a,u}^{dl}}\boldsymbol{h}_{u,a}^*\boldsymbol{f}_{a,u}s_u^{dl}}_{\text{Signal of interest}} 
\\

& + \underbrace{\displaystyle \sum_{i \in\mathbb{U}\setminus u} \displaystyle \sum_{a \in\mathbb{A}_i}\sqrt{\alpha_{a,i}^{dl}}\boldsymbol{h}_{u,a}^*\boldsymbol{f}_{a,i}s_i^{dl}}_{\text{Interference}}+\underbrace{w_u}_{\text{AWGN}},
\end{array}
\end{equation} 
where the desired signal at user $u$ is expressed by the first term, the interference components are represented by the second term, and the AWGN is described by the third term.

\subsection{Uplink Data Transmission}

In the uplink of the considered scenario, the users, both GUs and UAVs, are transmitting to the APs in the network. The received signal at the $a^{th}$ AP, $\boldsymbol{y}_a \in \mathbb{C}^{M_{AP} \times 1}$, is given by 
\begin{equation}
\label{eq:eq-2}
\begin{array}{ll}
\boldsymbol{y}_a& = \displaystyle \sum_{i \in\mathbb{U}} \sqrt{\alpha_i^{ul}}\boldsymbol{h}_{i,a}s_{i}^{ul}+\boldsymbol{w}_a,
\end{array}
\end{equation} 
where $\alpha_i^{ul}$ is the uplink power control coefficient of user $i$, $s_i^{ul}$ is the uplink data symbol to be transmitted by user $i$, and $\boldsymbol{w}_a \in \mathbb{C}^{M_{AP}\times 1}$ is the vector of the i.i.d. additive white Gaussian noise (AWGN) components at the $a^{th}$ AP. Thus, $ \boldsymbol{w}_a\sim \mathcal{CN}(0,\sigma_w^2\boldsymbol{I}_{M_{AP}}) $ with $\sigma_w^2$ is the corresponding noise variance. Equation (\ref{eq:eq-2}) can be rewritten as follows
\begin{equation}
\label{eq:eq-3}
\begin{array}{ll}
\boldsymbol{y}_a& = \sqrt{\alpha_u^{ul}}\boldsymbol{h}_{u,a}s_{u}^{ul}+ \displaystyle \sum_{i \in\mathbb{U} \setminus u} \sqrt{\alpha_i^{ul}}\boldsymbol{h}_{i,a}s_{i}^{ul}+\boldsymbol{w}_a,
\end{array}
\end{equation} 
where the first term is the desired signal of user $u$ at AP $a$, the second term is the corresponding interference term on the signal of user $u$, and the third term accounts for the AWGN. Every AP $a$ decodes the received vector for only the set of its served users, i.e. $\forall u \in \mathbb{U}_a$, by multiplying the received signal in (\ref{eq:eq-3}) by the uplink receive combining vector, $\boldsymbol{f}_{u,a}$ to obtain an estimate of the corresponding uplink data symbol, $\hat{s}_{u,a}^{ul}$, based on the local knowledge of AP $a$ as follows
\begin{equation}
\label{eq:eq-4}
\begin{array}{ll}
\hat{s}_{u,a}^{ul}& = \boldsymbol{f}^{*}_{u,a} \boldsymbol{y}_a, \forall u \in \mathbb{U}_a\\
& =  \sqrt{\alpha_u^{ul}}\boldsymbol{f}^{*}_{u,a}\boldsymbol{h}_{u,a}s_{u}^{ul} +   \displaystyle \sum_{i \in\mathbb{U} \setminus u} \sqrt{\alpha_i^{ul}}\boldsymbol{f}^{*}_{u,a}\boldsymbol{h}_{i,a}s_{i}^{ul}
\\
&+ \boldsymbol{f}^{*}_{u,a}\boldsymbol{w}_a , \forall u \in \mathbb{U}_a
\end{array}
\end{equation} 
Then, every AP $a$ sends the estimated $\hat{s}_{u,a}^{ul} \forall u \in \mathbb{U}_a$ to the CPU which can calculate the final estimate of the uplink symbol $\hat{s}_{u}$ for every user $u \in \mathbb{U}$ as follows 
\begin{equation} 
\label{eq:eq-4_2}
\begin{array}{ll}
\hat{s}_{u}^{ul} & = \displaystyle \sum_{a \in\mathbb{A}_u}\hat{s}_{u,a}^{ul}, \forall u \in \mathbb{U}
\\

& = \sqrt{\alpha_u^{ul}} \displaystyle \sum_{a \in\mathbb{A}_u} \boldsymbol{f}^{*}_{u,a} \boldsymbol{h}_{u,a}s_{u}^{ul} +  \displaystyle \sum_{a \in\mathbb{A}_u} \boldsymbol{f}^{*}_{u,a}\boldsymbol{w}_a
\\
&+  \displaystyle \sum_{i \in\mathbb{U} \setminus u} \sqrt{\alpha_i^{ul}}\sum_{a \in\mathbb{A}_u} \boldsymbol{f}^{*}_{u,a}\boldsymbol{h}_{i,a}s_{i}^{ul}, \forall u \in \mathbb{U},
\end{array}
\end{equation}  
where the cumulative desired signal at the CPU is represented by the first term, the cumulative interference at the CPU is represented by the third term, and the total  AWGN at the CPU is represented by the second term in (\ref{eq:eq-4_2}).

\section{Problem Formulation} \label{sec:formulation}

In this section, we discuss the formulations of the two considered optimization problems, namely, the sum rate maximization and the minimum rate maximization for both the downlink and uplink of CFmMIMO network. From equation (\ref{eq:eq-4_DL}), we can conclude that the desired signal power at user $u$ is $\left | \displaystyle \sum_{a \in\mathbb{A}_u}\sqrt{\alpha_{a,u}^{dl}}\boldsymbol{h}_{u,a}^*\boldsymbol{f}_{a,u}\right |^2$ and the corresponding interference power is $\displaystyle \sum_{i \in\mathbb{U}\setminus u} \left | \displaystyle \sum_{a \in\mathbb{A}_i}\sqrt{\alpha_{a,i}^{dl}}\boldsymbol{h}_{u,a}^*\boldsymbol{f}_{a,i}\right |^2$. Thus, the downlink signal to interference plus noise ratio (SINR) at the user $u$, $\xi_u^{dl}$, is given by
\begin{equation}
\label{eq:eq-1_PF}
\begin{array}{ll}
\xi_u^{dl}& = \displaystyle \frac{\left | \displaystyle \sum_{a \in\mathbb{A}_u}\sqrt{\alpha_{a,u}^{dl}}\boldsymbol{h}_{u,a}^*\boldsymbol{f}_{a,u}\right |^2}{ \displaystyle \sum_{i \in\mathbb{U}\setminus u} \left | \displaystyle \sum_{a \in\mathbb{A}_i}\sqrt{\alpha_{a,i}^{dl}}\boldsymbol{h}_{u,a}^*\boldsymbol{f}_{a,i}\right |^2+\sigma_u^2}. 
\end{array}
\end{equation}
Practically, due to the unavailability of the actual channel vectors $\boldsymbol{h}_{u,a}$ and to alleviate the CSI exchanging for the power control optimization, we define the practically obtainable SINR, $ \gamma_u^{dl}$, which can be calculated based on $\boldsymbol{\hat{h}}_{LS;u,a}$ as follows
\begin{equation}
\label{eq:eq-1_PF_2}
\begin{array}{ll}
\gamma_u^{dl}& = \displaystyle \frac{\left | \displaystyle \sum_{a \in\mathbb{A}_u}\sqrt{\alpha_{a,u}^{dl}}\boldsymbol{\hat{h}}_{LS;u,a}^*\boldsymbol{f}_{a,u}\right |^2}{ \displaystyle \sum_{i \in\mathbb{U}\setminus u} \left | \displaystyle \sum_{a \in\mathbb{A}_i}\sqrt{\alpha_{a,i}^{dl}}\boldsymbol{\hat{h}}_{LS;u,a}^*\boldsymbol{f}_{a,i}\right |^2+\sigma_u^2}. 
\end{array}
\end{equation} 
For the sake of simplifying the downlink SINR expression while not ignoring that fact that the distance from a certain user to different APs can differ, we assume that  $ \alpha_{a,u}^{dl} = b_{u,a}\alpha_{u}^{dl} \forall a \in \mathbb{A}_u $. Here, $\alpha_{u}^{dl}$ denotes the total downlink power allocated in the downlink from the set of APs $ \mathbb{A}_u$ serving user $u$ for certain user, and $b_{u,a}$ determines the fraction of the total power $\alpha_{u}^{dl}$ to be transmitted by AP $a \in \mathbb{A}_u$, that is given by
\begin{equation}
\label{eq:eq-1_PF_b}
\begin{array}{ll}
b_{u,a}&=\displaystyle \frac{\displaystyle \frac{1}{\beta_{u,a}}}{\displaystyle \sum_{\acute{a}\in \mathbb{A}_u}\displaystyle \frac{1}{\beta_{u,\acute{a}}}}, \forall a \in \mathbb{A}_u.
\end{array}
\end{equation}
Thus equation (\ref{eq:eq-1_PF_2}) can be rewritten as 
\begin{equation}
\label{eq:eq-2_PF}
\begin{array}{ll}
\gamma_u^{dl}& = \displaystyle \frac{\alpha_{u}^{dl}\left | \displaystyle \sum_{a \in\mathbb{A}_u}\sqrt{b_{u,a}}\boldsymbol{\hat{h}}_{LS;u,a}^*\boldsymbol{f}_{a,u}\right |^2}{ \displaystyle \sum_{i \in\mathbb{U}\setminus u} \alpha_{i}^{dl}\left | \displaystyle \sum_{a \in\mathbb{A}_i}\sqrt{b_{i,a}}\boldsymbol{\hat{h}}_{LS;u,a}^*\boldsymbol{f}_{a,i}\right |^2+\sigma_u^2}.
\end{array}
\end{equation}
On the other hand, from equation (\ref{eq:eq-4_2}), it is clear that the practically obtainable uplink SINR at the CPU level for the $u^{th}$ user, $\gamma_u^{ul}$, is given by equation (\ref{eq:eq-3_PF}). 
\begin{equation}
\label{eq:eq-3_PF} 
\begin{array}{ll}
\gamma_u^{ul}& = \displaystyle \frac{\alpha_u^{ul} \left | \displaystyle \sum_{a \in\mathbb{A}_u}\boldsymbol{f}_{u,a}^* \boldsymbol{\hat{h}}_{LS;u,a}\right |^2}{ \displaystyle \sum_{i \in\mathbb{U}\setminus u} \alpha_i^{ul} \displaystyle  \left | \sum_{a \in\mathbb{A}_u}\boldsymbol{f}_{u,a}^* \boldsymbol{\hat{h}}_{LS;i,a}\right |^2+\sigma_w^2 \sum_{a \in\mathbb{A}_u} \left\| \boldsymbol{f}^{*}_{u,a}\right\|^2}.
\end{array}
\end{equation}
We can use a general expression for the SINR that can represent both uplink and downlink SINR equations (\ref{eq:eq-2_PF}) and (\ref{eq:eq-3_PF}) as follows
\begin{equation}
\label{eq:eq-4_PF}
\begin{array}{ll}
\gamma_u& = \displaystyle \frac{\alpha_u q_u}{ \displaystyle \sum_{i \in\mathbb{U}\setminus u} \alpha_i e_{u,i}+t_{u}},
\end{array}
\end{equation}
where $\alpha_u$ is expressed as
\begin{equation}
\label{eq:eq-5_1_PF}
\alpha_u =  \left\{\begin{array}{l}
\alpha_u^{dl}, \textnormal{for downlink},\\ 
\alpha_u^{ul}, \textnormal{for  uplink},
\end{array}\right.
\end{equation}
$q_u$ is given by 
\begin{equation}
\label{eq:eq-5_PF}
q_u =  \left\{\begin{array}{l}
\left | \displaystyle \sum_{a \in\mathbb{A}_u}\sqrt{b_{u,a}}\boldsymbol{\hat{h}}_{LS;u,a}^*\boldsymbol{f}_{a,u}\right |^2, \textnormal{for downlink},\\ 
\left | \displaystyle \sum_{a \in\mathbb{A}_u}\boldsymbol{f}_{u,a}^* \boldsymbol{\hat{h}}_{LS;u,a}\right |^2, \textnormal{for  uplink},
\end{array}\right.
\end{equation}
 $e_{u,i}$ is represented by
\begin{equation}
\label{eq:eq-6_PF}
e_{u,i} = \left\{\begin{array}{l}
\left | \displaystyle \sum_{a \in\mathbb{A}_i}\sqrt{b_{i,a}}\boldsymbol{\hat{h}}_{LS;u,a}^*\boldsymbol{f}_{a,i}\right |^2, \textnormal{for downlink},\\ 
 \left | \displaystyle \sum_{a \in\mathbb{A}_u}\boldsymbol{f}_{u,a}^* \boldsymbol{\hat{h}}_{LS;i,a}\right |^2, \textnormal{for  uplink},
\end{array}\right.
\end{equation}
and $t_{u}$ is expressed by
\begin{equation}
\label{eq:eq-7_PF}
t_{u} = \left\{\begin{array}{ll}
\sigma_u^2, &\textnormal{for downlink},\\ 
 \sigma_w^2 \sum_{a \in\mathbb{A}_u} \left\| \boldsymbol{f}^{*}_{u,a}\right\|^2, &\textnormal{for  uplink}.
\end{array}\right.
\end{equation}

In the sequel of the problem formulation and the explanation of the proposed schemes, we will use the general form of the SINR presented in equation (\ref{eq:eq-4_PF}) where the discussions applies for both the uplink and the downlink. Thus, in the FBLC regime, the corresponding achievable data rate of the user $u$ can be expressed as in \cite{polyanskiy2010channel, mary2016finite, nasir2020resource}, wherein the interference can be treated as Gaussian noise \cite{yi2019opportunistic, ostman2021urllc}
\begin{equation}
\label{eq:eq-8_PF}
\begin{array}{ll}
R_u& = \displaystyle \frac{\tau_c-\tau_p}{2}B \left ( \vphantom{\displaystyle \frac{Q^{-1}(\epsilon)}{\sqrt{tB}}}  {\log}_2 (1+ \gamma_u) - \left( 1-\frac{1}{(1+ \gamma_u)^2}\right)^{1/2} \right.
\\  
 & \left. \times \displaystyle \frac{Q^{-1}(\epsilon)}{\sqrt{TB}}\log_2 \left(e \right) \right ) \quad{\mbox{bits/sec}}, 
\end{array}
\end{equation} 
where $B$ is the employed bandwidth, $\epsilon$ is the block error probability, $e$ is the natural exponent, $T$ is the transmission duration, and $Q^{-1}(.)$ is the inverse of the Q-function, which is the complementary cumulative distribution function of the standard Gaussian random variable. 

The first term in equation (\ref{eq:eq-8_PF}) is the traditional Shannon's data rate, $R_u^{sh}$, which is expressed as  
\begin{equation}
\label{eq:eq-8_PF_sh}
\begin{array}{ll}
R_{u}^{sh}& = \displaystyle \frac{\tau_c-\tau_p}{2}B \  \log_2 (1+ \gamma_u) = c_1 \  \log_2 (1+ \gamma_u),
\end{array}
\end{equation} 
where $c_1$ is a constant that replaces $\frac{\tau_c-\tau_p}{2}B$. On the other hand, the second term in (\ref{eq:eq-8_PF}) corresponds to the channel dispersion which can be regarded as a correction factor due to the FBLC regime, where the transmission duration of every individual transmitted packet is assumed to be much less than the coherence time of the communication channel. This correction term is denoted by $\Delta_u$ and is given by
\begin{equation}
\label{eq:eq-8_PF_FBLC_correction}
\begin{array}{ll}
\Delta_u &= c_1 \left( 1-\frac{1}{(1+ \gamma_u)^2}\right)^{1/2}   
  \displaystyle \frac{Q^{-1}(\epsilon)}{\sqrt{TB}}\log_2 \left(e \right) 
  \\
  &= c_2 \left( 1-\frac{1}{(1+ \gamma_u)^2}\right)^{1/2},
\end{array}
\end{equation} 
where the constant $c_2$ equals $ c_1 \frac{Q^{-1}(\epsilon)}{\sqrt{TB}}\log_2 \left(e \right)$. Thus, the FBLC data rate expression in (\ref{eq:eq-8_PF}) can be rewritten as 
\begin{equation}
\label{eq:eq-8_PF_new}
\begin{array}{ll}
R_u&= R_{u}^{sh} - \Delta_u = c_1 \log_2 (1+ \gamma_u) - c_2 \left( 1-\frac{1}{(1+ \gamma_u)^2}\right)^{1/2}.
\end{array}
\end{equation} 

In this work, we consider two distinct optimization problems. The objective of the first one is to find the set of the power control coefficients $\alpha_{u}, \forall u$ such that the total sum rate of all served users is maximized as presented in (\ref{eq:eq-10_PF})
\begin{equation}
\label{eq:eq-10_PF}
\begin{array}{cl}
\underset{\alpha_{u}, \forall u}{\textnormal{max}} & \displaystyle \sum_{u \in\mathbb{U}} R_u.
\end{array}
\end{equation}
The second optimization problem's objective is to find the set of the power control coefficients $\alpha_{u}, \forall u$ so as to maximize the minimum user's rate as in (\ref{eq:eq-11_PF})
\begin{equation}
\label{eq:eq-11_PF}
\begin{array}{cl}
\underset{\alpha_{u}, \forall u}{\textnormal{max}} & \displaystyle \min_{u \in\mathbb{U}} R_u.
\end{array}
\end{equation} 
The objectives (\ref{eq:eq-10_PF}) and (\ref{eq:eq-11_PF}) will be separately optimized for both the uplink and the downlink. In the case of the uplink, a constraint on the transmit power of every end user's device either a GU or an UAV will be imposed as follows
\begin{equation}
\label{eq:eq-12_PF}
\begin{array}{cl}
 & 0\leq \alpha_{u} \leq \eta_u, \forall u\in \mathbb{U},
\end{array}
\end{equation}
where $\eta_u$ is the maximum allowable power that can be transmitted by user $u$ in the uplink. For the downlink, and taking into account that the transmit power of AP $a$ to user $u$  is $ \alpha_{a,u}^{dl} =b_{u,a} \alpha_{u}=b_{u,a} \alpha_{u}^{dl}$, a constraint on the power budget of every AP in the network should be fulfilled
\begin{equation}
\label{eq:eq-13_PF}
\begin{array}{cl}
 & 0\leq \displaystyle \sum_{u \in\mathbb{U}_a}b_{u,a}\alpha_{u} \leq \eta_a, \forall a\in \mathbb{A},
\end{array}
\end{equation}
where $\eta_a$ is the maximum allowable power budget for AP $a$.

Thus, the sum rate maximization problem for the downlink is formulated as follows
\begin{equation}
\label{eq:eq-14_PF} 
\begin{array}{cl}
 \textnormal{(\ref{eq:eq-10_PF})}& \textnormal{s.t. (\ref{eq:eq-13_PF})},
\end{array}
\end{equation}
and the minimum rate maximization problem for the downlink is presented as follows
\begin{equation}
\label{eq:eq-15_PF}
\begin{array}{cl}
 \textnormal{(\ref{eq:eq-11_PF})}& \textnormal{s.t. (\ref{eq:eq-13_PF})}.
\end{array}
\end{equation}
On the other hand, the sum rate maximization problem for the uplink is given by
\begin{equation}
\label{eq:eq-16_PF}
\begin{array}{cl}
 \textnormal{(\ref{eq:eq-10_PF})}& \textnormal{s.t. (\ref{eq:eq-12_PF})},
\end{array}
\end{equation}
and the minimum rate maximization problem for the downlink is presented as follows
\begin{equation}
\label{eq:eq-17_PF}
\begin{array}{cl}
 \textnormal{(\ref{eq:eq-11_PF})}& \textnormal{s.t. (\ref{eq:eq-12_PF})}.
\end{array}
\end{equation}

\section{Proposed URLLC Rate Approximation Schemes for Successive Convex optimization}\label{sec:proposed}
The four considered optimization problems (\ref{eq:eq-14_PF}), (\ref{eq:eq-15_PF}), (\ref{eq:eq-16_PF}), and (\ref{eq:eq-17_PF}) are all neither convex nor concave. The constaints in all of these four problems are linear inequalities; however, the issue is with the individual users' URLLC rate expression (\ref{eq:eq-8_PF_new}) which consists of two terms that are nonconvex and nonconcave. Thus, we present two schemes for approximating this URLLC rate expression. The two schemes are applicable when optimizing both the total sum rate and the minimum user's  rate in both the uplink and the downlink. Specifically, we present two concave approximations for the data rate expression (\ref{eq:eq-8_PF_new}) to circumvent the fact that the original FBLC data rate expression (\ref{eq:eq-8_PF_new}) is both complicated and nonconcave. Thus, we resort to an iterative algorithm based on successive convex optimization (SCO). The basic idea of the two proposed schemes is to take a feasible power control coefficient vector obtained by solving the considered optimization problem in the previous iterative step as the starting point. In other words, at the $k^{th}$ iterative step, we take a starting feasible point $\boldsymbol{\alpha}^{(k-1)}= [\alpha_1^{(k-1)},\alpha_2^{(k-1)},..,\alpha_U^{(k-1)}]^'$, where $(.)^'$ represents the vector transpose operation. Then, at this feasible point, we find an approximate concave representation for the URLLC expression (\ref{eq:eq-8_PF_new}) by applying any of the two proposed approximations. This renders the considered optimization problems convex. The obtained new vector $\boldsymbol{\alpha}^{(k)}$ will be used as the input to the next iteration, i.e., $(k+1)^{th}$ iteration. This should be repeated until the convergence of $\boldsymbol{\alpha}$. The considered convergence criterion is given by
\begin{equation}
\label{eq:RA-1}
\begin{array}{ll}
\displaystyle\frac{\|\boldsymbol{\alpha}^{(k)}-\boldsymbol{\alpha}^{(k-1)}\|^2}{\|\boldsymbol{\alpha}^{(k)}\|^2} \leq \delta,
\end{array}
\end{equation}
where $\delta$ is a small acceptable tolerance for the convergence of $\boldsymbol{\alpha}^{(k)}$.

\subsection{First Scheme: Successive Convex Optimization with Iterative Concave Lower Bound Approximation} 
In this part, we explain the details of the first proposed scheme which is based on successive convex optimization with iterative concave lower bound approximation (SCO-ICBA). Our aim in this scheme is to find a concave lower bound for the URLLC rate expression in (\ref{eq:eq-8_PF_new}). Here we resort to successive convex approximation technique, where the URLLC rate function is approximated to a local tight approximate concave lower bound \cite{razaviyayn2014successive,scutari2013decomposition}. Specifically, we seek a concave lower bound for $ R_u^{sh}$ in equation (\ref{eq:eq-8_PF_sh}) and a linear upper bound for $\Delta_u$ in equation (\ref{eq:eq-8_PF_FBLC_correction}). So, the approximation for the URLLC data rate is given by
\begin{equation}
\label{eq:RA-ICBA_1}
\begin{array}{ll}
R_u^{ICBA}&= R_{u}^{sh,ICBA} - \Delta_u^{ICBA},
\end{array}
\end{equation}
where the $R_{u}^{sh,ICBA}$ is the concave lower bound approximation for the Shannon rate $R_{u}^{sh}$ expression in (\ref{eq:eq-8_PF_sh}). To find a concave lower bound, we use the inequality (\ref{eq:RA-ICBA_2_1}) \cite{nasir2021cell, sheng2018power}.
\begin{equation}
\label{eq:RA-ICBA_2_1}
\begin{array}{ll}
ln \left(1+\displaystyle \frac{x}{y}\right)\geq ln\left(1+\displaystyle \frac{\bar{x}}{\bar{y}}\right)+\displaystyle \frac{\bar{x}}{\bar{y}}\left(2\displaystyle\frac{\sqrt{x}}{\sqrt{\bar{x}}}-\displaystyle \frac{x+y}{\bar{x}+\bar{y}}-1\right),
\end{array}
\end{equation}
where we substitute $x$ with $\alpha_u q_u$ and $y$ with $\displaystyle \sum_{i \in\mathbb{U}\setminus u} \alpha_i e_{u,i}+t_{u}$. So, the concave lower bound for $R_{u}^{sh}$ can be expressed by (\ref{eq:RA-ICBA_2}) at the top of this page.
\begin{figure*}[!htb]
\begin{equation}
\label{eq:RA-ICBA_2}
\begin{array}{ll}
 R_{u}^{sh,ICBA}\approx R_{u}^{sh,ICBA(k)} & = \displaystyle\frac{c_1}{ln(2)} ln \left(1+ \displaystyle \frac{\alpha_u^{(k-1)} q_u}{ \displaystyle \sum_{i \in\mathbb{U}\setminus u} \alpha_i^{(k-1)} e_{u,i}+t_{u}}\right) 
+\displaystyle\frac{c_1}{ln(2)}\displaystyle \frac{\alpha_u^{(k-1)} q_u}{ \displaystyle \sum_{i \in\mathbb{U}\setminus u} \alpha_i^{(k-1)} e_{u,i}+t_{u}} \\
 &  \times \left (2\displaystyle \frac{\sqrt{\alpha_u}}{\sqrt{\alpha_u^{(k-1)}}}-\displaystyle \frac{\alpha_u q_u+\displaystyle \sum_{i \in\mathbb{U}\setminus u} \alpha_i e_{u,i}+t_{u}}{ \alpha_u^{(k-1)} q_u+\displaystyle \sum_{i \in\mathbb{U}\setminus u} \alpha_i^{(k-1)} e_{u,i}+t_{u}} -1\right)\leq R_{u}^{sh}.
\end{array}
\end{equation}
\end{figure*} 

While $\Delta_u$ defined by (\ref{eq:eq-8_PF_FBLC_correction}) is a nonconvex function in the power control coefficients $\alpha_u$, it is a convex function in the $\gamma_u$. In order to deal with $\Delta_u^{ICBA} $, let us find the Taylor expansion \cite{bronshtein2015handbook, attenborough2003mathematics} in terms of $\gamma_u$ as declared by (\ref{eq:RA-ICBA_3}). 
\begin{equation}
\label{eq:RA-ICBA_3}
\begin{array}{ll}
\Delta_u^{ICBA(k)} &\approx \Delta_u^{IIA}|_{\gamma_u=\gamma_u^{\left (  k \right )}}+ \displaystyle\frac{\partial \Delta_u^{IIA}}{\partial \gamma_u}|_{\gamma_u=\gamma_u^{\left (  k \right )}} \times \left(\gamma_u-\gamma_u^{\left (  k \right )}\right).
\end{array}
\end{equation}
Here $\gamma_u^{\left (  k \right )}$ is obtained from (\ref{eq:eq-4_PF})  by substituting $ \alpha_u, \text{ and } \alpha_i \forall u,i\in \mathbb{U}$ with $ \alpha_u^{\left (  k-1 \right )}, \text{ and } \alpha_i^{\left (  k-1 \right )}$, respectively, as follows 
\begin{equation}
\label{eq:RA-ICBA_4}
\begin{array}{ll}
\gamma_u^{\left (  k \right )}=\displaystyle \frac{\alpha_u^{\left (  k-1 \right )} q_u}{ \displaystyle \sum_{i \in\mathbb{U}\setminus u} \alpha_i^{\left (  k-1 \right )} e_{u,i}+t_{u}}.
\end{array}
\end{equation}
Now, the only part of equation (\ref{eq:RA-ICBA_3}) that contains the variables $\alpha_u, \text{ and } \alpha_i$ is $\gamma_u$ which is not convex as shown in (\ref{eq:eq-4_PF}). Our approach to deal with the fact that $\gamma_u$ is a nonconvex function with respect to the power control coefficients depends on the following inequality
\begin{equation}
\label{eq:RA-ICBA_5_1}
\begin{array}{ll}
x &\leq \displaystyle \frac{1}{2}\left(\displaystyle \frac{x^2}{\bar{x}}+\bar{x}\right).
\end{array}
\end{equation}
Specifically, we apply the inequality (\ref{eq:RA-ICBA_5_1}) on the variable $\alpha_u$ in the $\gamma_u$  formula (\ref{eq:eq-4_PF}). Thus, we can find a convex upper bound for $\gamma_u$ as presented by (\ref{eq:RA-ICBA_5}) at the top of the next page.
\begin{figure*}[!tb]
\begin{equation}
\label{eq:RA-ICBA_5}
\begin{array}{ll}
\gamma_u^{ICBA}\approx\gamma_u^{ICBA(k)}& = \displaystyle \frac{0.5\left(\displaystyle \frac{\alpha_u^2}{\alpha_u^{(k-1)}}+\alpha_u^{(k-1)}\right) q_u}{ \displaystyle \sum_{i \in\mathbb{U}\setminus u} \alpha_i e_{u,i}+t_{u}}
\\ &=\displaystyle \frac{\alpha_u^2q_u}{2\alpha_u^{(k-1)}\left(\displaystyle \sum_{i \in\mathbb{U}\setminus u} \alpha_i e_{u,i}+t_{u}\right)}
+\displaystyle \frac{\alpha_u^{(k-1)}q_u}{2\left(\displaystyle \sum_{i \in\mathbb{U}\setminus u} \alpha_i e_{u,i}+t_{u}\right)}\geq\gamma_u,
\end{array}
\end{equation}
\end{figure*}
 The upper bound in (\ref{eq:RA-ICBA_5}) is convex because it is a summation of two convex functions. So, we plug the upper bound (\ref{eq:RA-ICBA_5}) into (\ref{eq:RA-ICBA_3}). Then, we substitute both $ R_u^{sh,ICBA(k)}$ and $\Delta_u^{ICBA(k)}$ from (\ref{eq:RA-ICBA_2}) and (\ref{eq:RA-ICBA_3}) into (\ref{eq:RA-ICBA_1}). The resultant expression $ R_u^{ICBA(k)}$ is a concave lower bound of $R_u$. Thus, we substitute $R_u$ by the obtained lower $ R_u^{ICBA(k)}$ in the objectives (\ref{eq:eq-10_PF}) and (\ref{eq:eq-11_PF}) making the four considered optimization problems convex. By solving each of these four optimization problems for finding the optimal power control coefficients vector $\boldsymbol{\alpha}$. The obtained optimal solution in the $k^{th}$ iteration will be denoted by $\boldsymbol{\alpha}^{(k)}$, that is dealt as the starting point for the next iterative step $k+1$. This procedure will be repeated until  the vector $\boldsymbol{\alpha}$ converges. Algorithm \ref{alg:ICBA} presents the steps of the SCO-ICBA scheme.

\begin{algorithm}[!t]
    \SetAlgoLined
    \textbf{Initialization}: Start from an initial feasible point $\boldsymbol{\alpha}^{(0)}$, set $k=1$\;
    \While{(\ref{eq:RA-1}) is not satisfied and the maximum number of iterations is not reached}
    {
        Utilize the approximation $ R_u^{ICBA(k)}$ in the considered optimization problem\;
        Solve the obtained convex optimization problem to get $\boldsymbol{\alpha}^{(k)}$\;
        Set $k \leftarrow k+1$;
    }
\caption{SCO-ICBA algorithm} 
\label{alg:ICBA}
\end{algorithm}

\subsection{Second Scheme: Successive Convex Optimization with Iterative Interference Approximation} 
In this part, we elaborate on our second proposed scheme which is based on successive convex optimization with iterative interference approximation (SCO-IIA). The basic idea is to insert the iterative values of the $\alpha_i \forall i\in \mathbb{U}\setminus u$ in the interference term, i.e., the denominator term of the $\gamma_u$ in equation (\ref{eq:eq-4_PF}). So, the URLLC data rate is approximated by 
\begin{equation}
\label{eq:RA-IIA_1}
\begin{array}{ll}
R_u^{IIA}&= R_{u}^{sh,IIA} - \Delta_u^{IIA},
\end{array}
\end{equation}
with
\begin{equation}
\label{eq:RA-IIA_2}
\begin{array}{ll}
 R_{u}^{sh,IIA}& \approx R_{u}^{sh,IIA(k)} = c_1 \log_2 (1+ \displaystyle \frac{\alpha_u q_u}{ \displaystyle \sum_{i \in\mathbb{U}\setminus u} \alpha_i^{(k-1)} e_{u,i}+t_{u}}),
\end{array}
\end{equation} 
and $\Delta_u^{IIA}$ can be approximated by $\Delta_u^{IIA(k)}$ which is given by 
\begin{equation}
\label{eq:RA-IIA_3}
\begin{array}{ll}
 \Delta_u^{IIA(k)}= c_2 \left( 1-\left(1+ \displaystyle \frac{\alpha_u q_u}{ \displaystyle \sum_{i \in\mathbb{U}\setminus u} \alpha_i^{(k-1)} e_{u,i}+t_{u}}\right)^{-2}\right)^{1/2}.
\end{array}
\end{equation}
Here,  $R_{u}^{sh,IIA}$, is a concave function of $\alpha_u$; while $\Delta_u^{IIA}$ is a nonconvex function of $\alpha_u$. So, we use the first order Taylor expansion \cite{bronshtein2015handbook, attenborough2003mathematics} to linearize this second term as in (\ref{eq:RA-IIA_4}). 
\begin{equation}
\label{eq:RA-IIA_4}
\begin{array}{ll}
\Delta_u^{IIA(k)} &\approx \Delta_u^{IIA}|_{\alpha_u=\alpha_u^{\left (  k-1 \right )}}+  \displaystyle\frac{\partial \Delta_u^{IIA}}{\partial \alpha_u}|_{\alpha_u=\alpha_u^{\left (  k-1 \right )}} 
\\
&\times \left(\alpha_u-\alpha_u^{\left (  k-1 \right )}\right).
\end{array}
\end{equation}
So, we can express the approximated data rate at the $k^{th}$ iteration, $R_u^{IIA}$, as in equation (\ref{eq:RA-IIA_5}) at the top of this page.  
\begin{figure*}[!tb]
\begin{equation}
\label{eq:RA-IIA_5}
\begin{array}{ll} 
R_u^{IIA(k)} &= c_1 \log_2 (1+ \displaystyle \frac{\alpha_u q_u}{ \displaystyle \sum_{i \in\mathbb{U}\setminus u} \alpha_i^{(k-1)} e_{u,i}+t_{u}}) 
-\Delta_u^{IIA}|_{\alpha_u=\alpha_u^{\left (  k-1 \right )}} 
-\displaystyle\frac{\partial \Delta_u^{IIA}}{\partial \alpha_u}|_{\alpha_u=\alpha_u^{\left (  k-1 \right )}} \times \left(\alpha_u-\alpha_u^{\left (  k-1 \right )}\right).
\end{array}
\end{equation} 
\end{figure*}

The approximated $R_u^{IIA(k)}$ in (\ref{eq:RA-IIA_5}) is concave because it is the difference between a concave function of $\alpha_u$ and a linear function of $\alpha_u$.
Hence, in the objectives (\ref{eq:eq-10_PF}) and (\ref{eq:eq-11_PF}), we can  replace the data rate expression (\ref{eq:eq-8_PF_new}) by the approximated expression  (\ref{eq:RA-IIA_5}). This, renders the four considered optimization problems convex.  The optimal solution in the $k^{th}$ iteration will be denoted by $\boldsymbol{\alpha}^{(k)}$, and will be used as the starting point in the next iterative step $k+1$. This procedure will be repeated until convergence of the vector $\boldsymbol{\alpha}$. Algorithm \ref{alg:IIA} declares the steps of the SCO-IIA scheme. 

\begin{algorithm}[t]

    \SetAlgoLined
    \textbf{Initialization}: Start from an initial feasible point $\boldsymbol{\alpha}^{(0)}$, set $k=1$\;
    \While{(\ref{eq:RA-1}) is not satisfied and the maximum number of iterations is not reached}
    {
        Utilize the approximation $ R_u^{IIA(k)}$ in the considered optimization problem\;
        Solve the obtained convex optimization problem to get $\boldsymbol{\alpha}^{(k)}$\;
        Set $k \leftarrow k+1$;
    }
\caption{SCO-IIA algorithm}
\label{alg:IIA}
\end{algorithm}

It is noteworthy that the initialization, i.e., selecting the initial value of the feasible point, $\boldsymbol{\alpha}^{(0)}= [\alpha_1^{(0)},\alpha_2^{(0)},..,\alpha_U^{(0)}]^'$, that will be used as the starting point in the first iteration, is very important and can impact the performance of the algorithm itself. For the sake of fair comparisons between the two considered schemes, we will use the same $\boldsymbol{\alpha}^{(0)}$ for both schemes. Specifically, we use the approximation $R_{u}^{sh,IIA}$ given in (\ref{eq:RA-IIA_2}) and plug it in the sum rate optimization problems, (\ref{eq:eq-14_PF}) for the downlink and  (\ref{eq:eq-16_PF}) for the uplink to replace the URLLC rate $R_{u}$. Then, we use the aforementioned procedure for iterating the corresponding sum rate maximization problem. Thus, we can obtain $\boldsymbol{\alpha}^{(0)}$ that will be used as the initialization for solving the corresponding optimization problems with both SCO-IIA and SCO-ICBA schemes. 

\section{Considered Beamforming Schemes}
We consider two distinct beamforming schemes which are used in both the uplink and the downlink of CFmMIMO network. In order to relieve the excessive CSI exchange, we will use the same beamforming vectors for all the $L$ coherence intervals that have the same large scale fading parameters. Specifically, we update the beamforming vectors every time the large scale fading changes, i.e., based on $\boldsymbol{\hat{h}}_{LS;u,a}$. First, we assume the PZF combining is employed at each AP for the uplink, and PZF transmission is used by the AP for the downlink data transmission in the considered CFmMIMO system. Specifically, for a certain user $u$ served by AP $a$, we are interested in the receive/transmit beamforming vector that can remove the strongest $N_I$ interfering signals, where $N_I<M_{AP}$. Here, the strengths of the interfering signals are measured in terms of the large scale fading coefficients of the corresponding channel vector. Note that, we are using the same frequency resources for all users served by the CFmMIMO network. For every user $u$ and at certain serving AP $a \in \mathbb{A}_u$, the set $\mathbb{E}_{u,a}$ of the estimated channel vectors from interfering users $i \in \mathbb{U} \setminus u$ is represented by 
\begin{equation}
\label{eq:eq-1_BF}
\begin{array}{ll}
 \mathbb{E}_{u,a}& := \{\boldsymbol{\hat{h}}_{LS;i,a}: i\in \mathbb{U} \setminus u\}
  \\
  & =\{\boldsymbol{\hat{h}}_{LS;1,a}, .., \boldsymbol{\hat{h}}_{LS;u-1,a}, \boldsymbol{\hat{h}}_{LS;u+1,a}, .., \boldsymbol{\hat{h}}_{LS;U,a}\}.
\end{array}
\end{equation} 
Then, the $\mathbb{E}_{u,a}$ will be sorted in descending order according to the large scale fading coefficients of every channel vector. We denote by $\mathbb{E}^O_{u,a}$, the set of the ordered interfering channels vectors which can be written as 
\begin{equation}
\label{eq:eq-2_BF}
\begin{array}{ll}
 \mathbb{E}^O_{u,a}& =\{\boldsymbol{\hat{h}}_{I(u,a)}^{O(1)}, \boldsymbol{\hat{h}}_{I(u,a)}^{O(2)}, .., \boldsymbol{\hat{h}}_{I(u,a)}^{O(U-1)}\},
\end{array}
\end{equation} 
where $\boldsymbol{\hat{h}}_{I(u,a)}^{O(k)}$ denotes the estimated $k^{th}$ strongest interfering channel vector on the channel between user $u$ and AP $a$ during the first coherence interval of every $L$ coherence intervals that have the same large scale fading parameters. So, the corresponding matrix of the $N_I$ strongest interfering channel vectors, $\boldsymbol{E}^I_{u,a}$, can be constructed as follows 
\begin{equation}
\label{eq:eq-3_BF}
\begin{array}{ll}
 \boldsymbol{E}^I_{u,a}& =[\boldsymbol{\hat{h}}_{I(u,a)}^{O(1)}, \boldsymbol{\hat{h}}_{I(u,a)}^{O(2)}, .., \boldsymbol{\hat{h}}_{I(u,a)}^{O(N_I)}].
\end{array}
\end{equation} 
Let $\boldsymbol{Q}_{u,a}$ denotes the matrix which columns are the orthonormal basis \cite{gilbert2016introduction} of the matrix $\boldsymbol{E}^I_{u,a}$. Define the vector $\boldsymbol{p}_{u,a}$ as 
\begin{equation}
\label{eq:eq-4_BF}
\begin{array}{ll}
 \boldsymbol{p}_{u,a}& =\left( \boldsymbol{I}-\boldsymbol{Q}_{u,a}\boldsymbol{Q}_{u,a}^*\right)\boldsymbol{\hat{h}}_{LS;u,a},
\end{array}
\end{equation} 
where $\boldsymbol{I}$ is the identity matrix. Thus, the employed PZF downlink transmission vector, $\boldsymbol{f}_{a,u}^{PZF}$, and PZF uplink combining vector, $\boldsymbol{f}_{u,a}^{PZF}$, are given by 
\begin{equation}
\label{eq:eq-5_BF}
\begin{array}{ll}
\boldsymbol{f}_{a,u}^{PZF} = \boldsymbol{f}_{u,a}^{PZF} = \displaystyle\frac{\boldsymbol{p}_{u,a}}{\left \|\boldsymbol{p}_{u,a} \right \|}.
\end{array}
\end{equation} 

Furthermore, we consider the MRT for the downlink and the MRC for the uplink, where corresponding downlink MRT vector, $\boldsymbol{f}_{a,u}^{MRT}$, and uplink MRT vector, $\boldsymbol{f}_{u,a}^{MRC}$, are given by 
\begin{equation}
\label{eq:eq-6_BF}
\begin{array}{ll}
\boldsymbol{f}_{a,u}^{MRT} = \boldsymbol{f}_{u,a}^{MRC} = \displaystyle\frac{\boldsymbol{\hat{h}}_{LS;u,a}}{\left \|\boldsymbol{\hat{h}}_{LS;u,a} \right \|}.
\end{array}
\end{equation} 
It is noteworthy that $\boldsymbol{f}_{a,u}^{PZF}$, $\boldsymbol{f}_{u,a}^{PZF}$, $\boldsymbol{f}_{a,u}^{MRT}$, and $\boldsymbol{f}_{u,a}^{MRC}$ are unit norm vectors which is consistent with our assumption in the system model section and the problem formulation section. Thus, substituting $\boldsymbol{f}_{a,u}$ by either $\boldsymbol{f}_{a,u}^{PZF}$ or $\boldsymbol{f}_{a,u}^{MRT}$, and $\boldsymbol{f}_{u,a}$ by either $\boldsymbol{f}_{u,a}^{PZF}$ or $\boldsymbol{f}_{u,a}^{MRC}$ will not require any modifications in the proposed schemes.
\section{Numerical Results}
In our experiments, we simulate a 1 km$^2$ square service area that is enwrapped around at its borders. The CFmMIMO system comprises $N_A = 100$ APs that are uniformly distributed through the considered service area, where each AP has a height of 10 m. Every AP has $M_{AP}=8$ antennas and it's maximum allowable transmit power is $\eta_a = 200$ mW. The channel bandwidth $B=20$ MHz with a center frequency $f=1.9$ GHz. The length of the coherence interval is assumed to be $\tau_c=200$ samples and the length of the uplink piolt trainig sequences is taken as $\tau_p=32$ samples. The URLLC transmission duration is assumed to be $T=5\times 10^{-5}$, while the target block error probability is taken by $\epsilon=10^{-5}$. We set the tolerance for the convergence of the power control coefficients vector to be $\delta=10^{-4}$ for both SCO-IIA and SCO-ICBA.

We consider the coexistence of 48 GUs along with 12 UAVs. The GUs are assumed to be uniformly distributed within the service area and each GU has a height of 1.65 m, while the UAVs are uniformly distributed within the service area at heights that range from 22.5 m to 300m. The uplink transmit power of every end user, either a GU or an UAVs, should not exceed $\eta_u = 100$ mW. We assume uplink pilot transmit power of every user $\alpha_{p,u}=\tau_p \eta_u = 3200$ mW. We assume that each receiver's thermal noise power spectral density, $N_0$, is $-174$ dBm/Hz, and the noise figure, $N_f$ of every receiver is 9 dB. Every user $u$, either a GU or an UAV, is served by the 5 APs that has the highest large scale fading coefficients to this user $u$, i.e.,  $N_{A,u}=5$. 

For the channel model of GUs, we assume an urban environment where the small scale fading is Rayleigh, i.e., the channel from any GU to an AP does not include a line of sight (LOS) component. The large scale fading coefficient between GU $g$ and AP $a$, $\beta_{g,a}$ in dB, is simulated as described in \cite{8952782, 3gpp2017further}
\begin{equation}
\label{eq:eq-19}
\begin{array}{l}
\beta_{g,a}=-36.7\log_{10}(d_{g,a})-22.7-26\log_{10}(f)+\zeta_{g,a},
\end{array}
\end{equation}
where $d_{g,a}$ denotes the distance from GU $g$ to AP $a$, $f$ denotes the center frequency in GHz, and $\zeta_{g,a} \sim \mathcal{N}(0,4^2)$ represents the impact of shadowing. The correlation between shadowing effects of GU $g$ and $i$ is represented as in \cite{8845768} 
\begin{equation}
\label{eq:eq-20}
\mathbb{E} \{\beta_{g,a},\beta_{i,b}\}=\left\{\begin{matrix}
4^2 2^{\frac{-r_{g,i}}{r_0}}, a=b,\\ 
0, \ \ \  \ \ \ \ \ a\neq b,
\end{matrix}\right.
\end{equation}
where $r_{g,i}$ is the distance from GU $u$ to GU $i$ and $r_0=9 m$.

For simulating the channel between an UAV and an AP, we consider urban-micro with aerial vehicles (UMi-AV) scenario \cite{muruganathan2018overview, meredith2017study}. Particularly, the LOS probability for a certain UAV is modeled as described in \cite[Table B-1]{meredith2017study}. Furthermore, we follow the specifications mentioned in \cite[Table B-2]{meredith2017study} to model the large scale fading between the UAVs and APs.

In our experiments, we evaluate the performance of the two proposed SCO-IIA and SCO-ICBA schemes for both the considered beamforming schemes, namely PZF reception/transmission and MRC/MRT, when applied on both the uplink and downlink of a CFmMIMO network. For the PZF case when applied to the CFmMIMO, we assume that $N_I=5$. For the purpose of benchmarking, the two proposed SCO-IIA and SCO-ICBA schemes are evaluated for generic ZF reception/transmission and MRC/MRT for both the uplink and downlink of traditional COmMIMO system. For simulating the COmMIMO system, we assume that the considered service area is served by 4 base stations, and each has 200 antennas. We consider that the the power that can be transmitted by every base station in the COmMIMO is $5$ W. 
Unless otherwise mentioned, we perform simulations for 250 scenarios with the considered simulation parameters; where we evaluate the performance of a single coherence interval. The performance of both SCO-IIA and SCO-ICBA scehmes under the considered beamforming techniques with both COmMIMO and CFmMOIMO for both the downlink and the uplink are evaluated in terms of the corresponding cumulative distribution function (CDF) plots of both GUs' rate and UAVs' rate.

\subsection{Performance of Sum Rate Maximization}
\begin{figure*}[!t]
 \centering
 \captionsetup{font=small}
\begin{subfigure}[b]{0.38\linewidth}
\centering
\captionsetup{font=small}
  \includegraphics[width=\columnwidth]{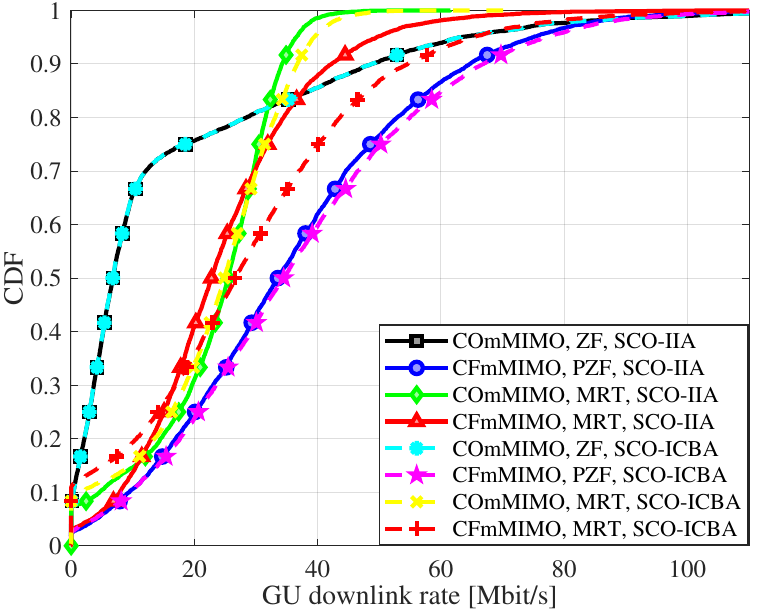} 
  \caption{GUs}
  \label{fig:RateCDF_DL_SumRate_GUs}
\end{subfigure}\hfil 
\begin{subfigure}[b]{0.38\linewidth}
\centering
\captionsetup{font=small}
  \includegraphics[width=\columnwidth]{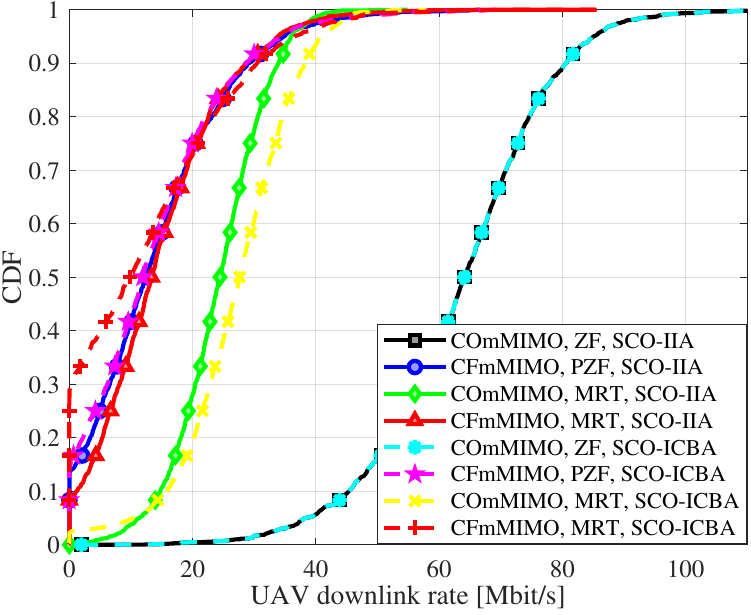}
  \caption{UAVs}
  \label{fig:RateCDF_DL_SumRate_UAVs}
\end{subfigure} 
\caption{Downlink FBLC rate comparison among the proposed SCO-IIA and SCO-ICBA when applied for sum rate maximization under COmMIMO with ZF beamforming, CFmMIMO with PZF beamforming, COmMIMO with MRT beamforming, and CFmMIMO with MRT beamforming.} 
\label{fig:RateCDF_DL_SumRate}
\end{figure*}
\begin{figure*}[!t]
 \centering
 \captionsetup{font=small}
\begin{subfigure}[b]{0.38\linewidth}
\centering
\captionsetup{font=small}
  \includegraphics[width=\columnwidth]{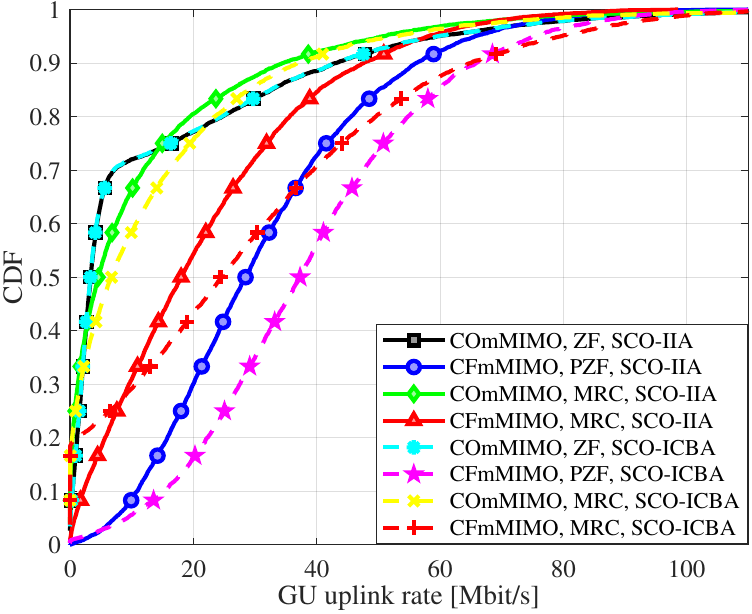} 
  \caption{GUs}
  \label{fig:RateCDF_UL_SumRate_GUs}
\end{subfigure}\hfil 
\begin{subfigure}[b]{0.38\linewidth}
\centering
\captionsetup{font=small}
  \includegraphics[width=\columnwidth]{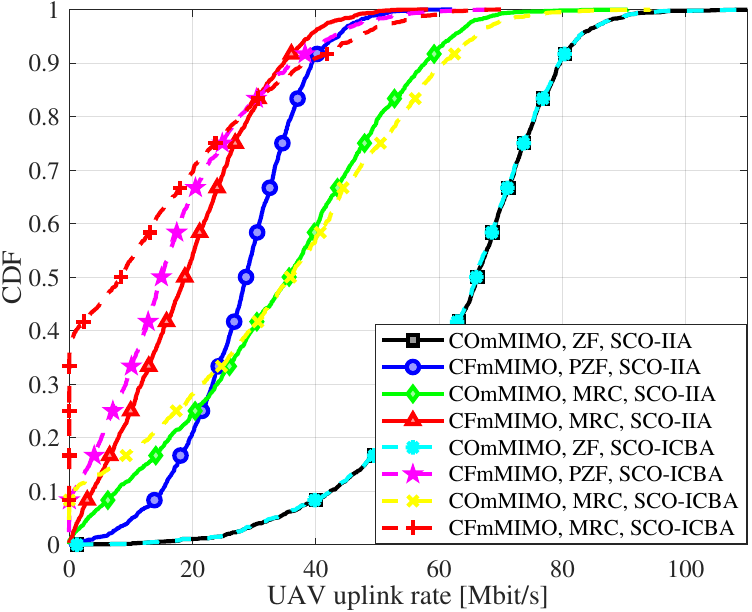}
  \caption{UAVs}
  \label{fig:RateCDF_UL_SumRate_UAVs}
\end{subfigure} 
\caption{Uplink FBLC rate comparison among the proposed SCO-IIA and SCO-ICBA when applied for sum rate maximization under COmMIMO with ZF reception, CFmMIMO with PZF reception, COmMIMO with MRC reception, and CFmMIMO with MRC reception.} 
\label{fig:RateCDF_UL_SumRate}
\end{figure*}

Fig. \ref{fig:RateCDF_DL_SumRate} declares the CDF plots of the downlink rates for GUs in Fig. \ref{fig:RateCDF_DL_SumRate_GUs} and  UAVs in Fig. \ref{fig:RateCDF_DL_SumRate_UAVs} with both COmMIMO and CFmMIMO under the considered beamforming techniques when optimizing the users' total sum rate considering both SCO-IIA and SCO-ICBA schemes. It is clear from Fig. \ref{fig:RateCDF_DL_SumRate_GUs} that, when the total system sum rate is the objective to be maximized, the GUs' downlink rate CDF performance of the CFmMIMO is better than its counterpart of the COmMIMO for both the considered SCO-IIA and SCO-ICBA schemes regardless of the employed beamforming technique except only for MRT with SCO-ICBA. Furthermore, in CFmMIMO, PZF  transmission can provide better GUs' downlink rate compared to MRT. In COmMIMO, the MRT outperforms the ZF transmission in terms of GUs' downlink rate. On the other hand, Fig. \ref{fig:RateCDF_DL_SumRate_UAVs} declares that COmMIMO provides better UAVs' rate performance compared to  CFmMIMO regardless of the utilized beamforming technique. Also, the UAVs' rate of COmMIMO with ZF transmission is much better that that of the COmMIMO with MRT. Note that, the downlink rate CDF plots of SCO-IIA scheme are very close to those of  SCO-ICBA scheme for both GUs' and UAVs' in all considered network-beamforming combinations except only for CFmMIMO with MRT where the performance of SCO-IIA is slightly better than that of SCO-ICBA. 

As for the uplink, Fig. \ref{fig:RateCDF_UL_SumRate} shows the CDF plots of the uplink rates for both GUs and  UAVs in Fig. \ref{fig:RateCDF_UL_SumRate_GUs} and Fig. \ref{fig:RateCDF_UL_SumRate_UAVs}, respectively, when served by COmMIMO and CFmMIMO under the considered receive beamforming techniques while maximizing the total sum rate with both SCO-IIA and SCO-ICBA schemes. Fig. \ref{fig:RateCDF_UL_SumRate_GUs} declares that, when optimizing the total system sum rate, the GUs' uplink rate CDF performance of the CFmMIMO is better than its counterpart of the COmMIMO for both the considered SCO-IIA and SCO-ICBA schemes. Moreover, in CFmMIMO, PZF  reception outperforms MRC in terms of GUs' uplink rate. In COmMIMO, the MRC has GUs' performance that is slightly better than that of the ZF reception. In addition, Fig. \ref{fig:RateCDF_UL_SumRate_UAVs} indicates that COmMIMO provides better UAVs' uplink rate performance compared to  CFmMIMO for both ZF and MRC beamforming techniques. Also, the UAVs' uplink rate of COmMIMO with ZF transmission is better than that of the COmMIMO with MRC. It is noteworthy that, the uplink rate CDF plots of SCO-IIA scheme are slightly better than those of  SCO-ICBA scheme for UAVs' in all considered network-beamforming scenarios except only for COmMIMO with ZF reception where the performance of SCO-ICBA is very close to that of SCO-IIA.

\begin{figure*}[!t]
 \centering
 \captionsetup{font=small}
\begin{subfigure}[b]{0.38\linewidth}
\centering
\captionsetup{font=small}
  \includegraphics[width=\columnwidth]{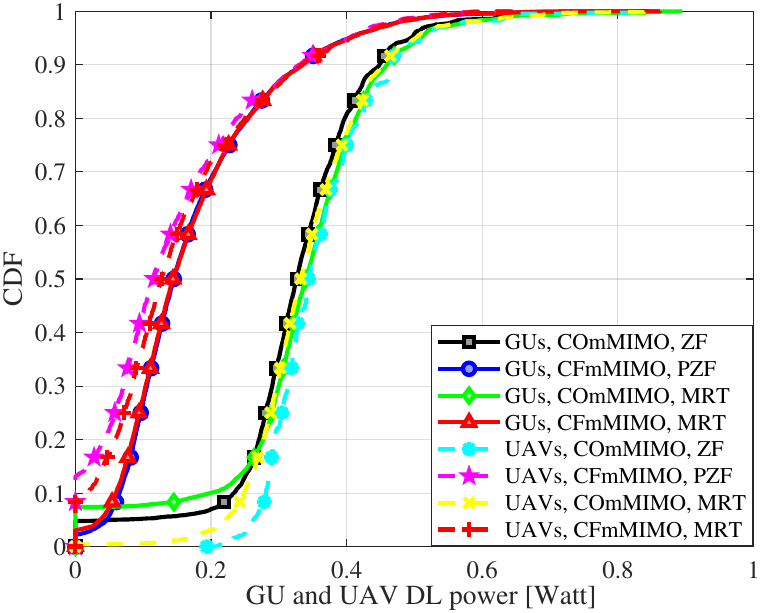} 
  \caption{SCO-IIA}
  \label{fig:PowerCDF_DL_SumRate_IIA}
\end{subfigure}\hfil 
\begin{subfigure}[b]{0.38\linewidth}
\centering
\captionsetup{font=small}
  \includegraphics[width=\columnwidth]{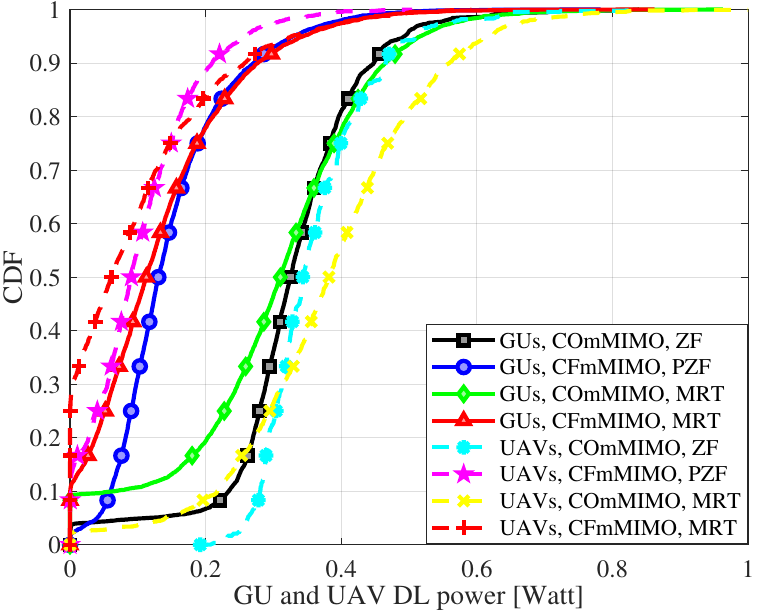}
  \caption{SCO-ICBA}
  \label{fig:PowerCDF_DL_SumRate_ICBA}
\end{subfigure} 
\caption{Downlink power consumption comparison between GUs and UAVs when both SCO-IIA and SCO-ICBA schemes are applied for sum rate maximization under COmMIMO with ZF transmission, CFmMIMO with PZF transmission, COmMIMO with MRT, and CFmMIMO with MRT.} \vspace{-0.5cm}
\label{fig:PowerCDF_DL_SumRate}
\end{figure*}
Fig. \ref{fig:PowerCDF_DL_SumRate} depicts the downlink power consumption comparison between GUs and UAVs when the sum rate is optimized using both SCO-IIA and SCO-ICBA schemes. Fig. \ref{fig:PowerCDF_DL_SumRate_IIA} and Fig. \ref{fig:PowerCDF_DL_SumRate_ICBA} declare that while maximizing the sum rate using either the SCO-IIA scheme or the SCO-SCA, the COmMIMO always consumes more power, compared to CFmMIMO, for both UAVs and GUs regardless of the the  beamforming scheme employed. When SCO-IIA is employed, the CDF plots of the power allocated to UAVs are closer to that of the GUs with all beamforming schemes compared to SCO-ICBA which means that  fairness of SCO-IIA is better than that of  SCO-ICBA if the sum rate is optimized. 

\subsection{Performance of Minimum Rate Maximization}
\begin{figure*}[!t]
 \centering
 \captionsetup{font=small}
\begin{subfigure}[b]{0.38\linewidth}
\centering
\captionsetup{font=small}
  \includegraphics[width=\columnwidth]{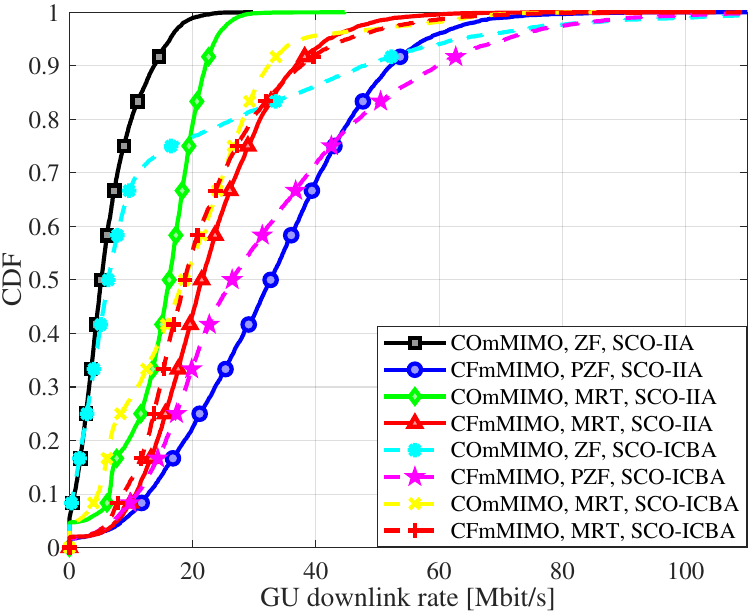} 
  \caption{GUs}
  \label{fig:RateCDF_DL_MinRate_GUs}
\end{subfigure}\hfil 
\begin{subfigure}[b]{0.38\linewidth}
\centering
\captionsetup{font=small}
  \includegraphics[width=\columnwidth]{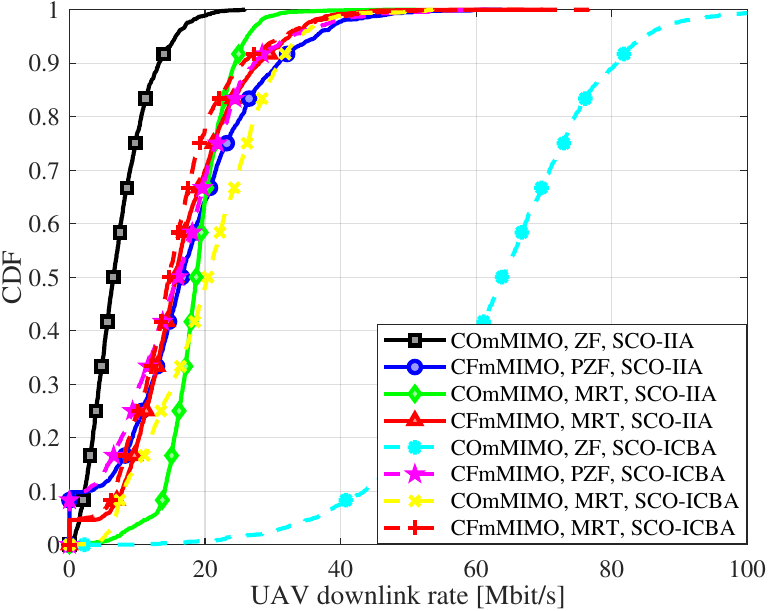}
  \caption{UAVs}
  \label{fig:RateCDF_DL_MinRate_UAVs}
\end{subfigure} 
\caption{Downlink FBLC rate comparison among the proposed SCO-IIA and SCO-ICBA when applied for minimum user's rate maximization under COmMIMO with ZF beamforming, CFmMIMO with PZF beamforming, COmMIMO with MRT beamforming, and CFmMIMO with MRT beamforming.} 
\label{fig:RateCDF_DL_MinRate}
\end{figure*}
 
\begin{figure*}[!t]
 \centering
 \captionsetup{font=small}
\begin{subfigure}[b]{0.38\linewidth}
\centering
\captionsetup{font=small}
  \includegraphics[width=\columnwidth]{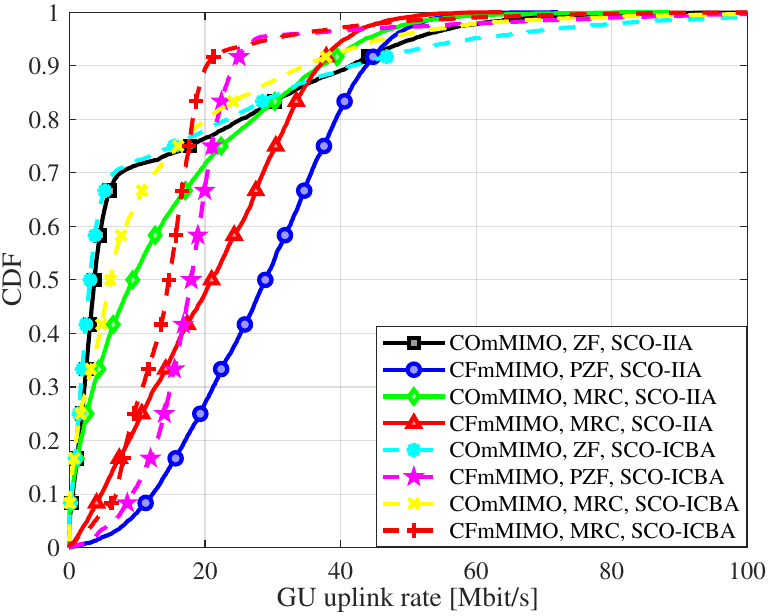} 
  \caption{GUs}
  \label{fig:RateCDF_UL_MinRate_GUs}
\end{subfigure}\hfil 
\begin{subfigure}[b]{0.38\linewidth}
\centering
\captionsetup{font=small}
  \includegraphics[width=\columnwidth]{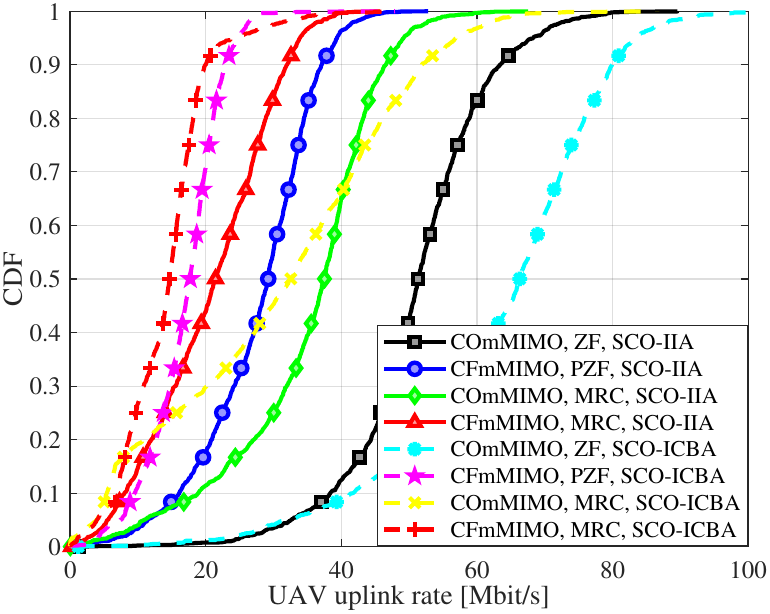}
  \caption{UAVs}
  \label{fig:RateCDF_UL_MinRate_UAVs}
\end{subfigure} 
\caption{Uplink FBLC rate comparison among the proposed SCO-IIA and SCO-ICBA when applied for minimum user's rate maximization under COmMIMO with ZF reception, CFmMIMO with PZF reception, COmMIMO with MRC reception, and CFmMIMO with MRC reception.} 
\label{fig:RateCDF_UL_MinRate}
\end{figure*}

Fig. \ref{fig:RateCDF_DL_MinRate} depicts the CDF plots of the downlink rates for both GUs in Fig. \ref{fig:RateCDF_DL_MinRate_GUs} and  UAVs in Fig. \ref{fig:RateCDF_DL_MinRate_UAVs} when served by COmMIMO and CFmMIMO under the considered beamforming techniques while optimizing the worst user's rate by using both SCO-IIA and SCO-ICBA schemes. Obviously, when maximizing the worst user's rate, the best GUs' downlink performance can be achieved using SCO-IIA scheme with CFmMIMO under PZF beamforming at the price of the UAVs performance as depicted from Fig. \ref{fig:RateCDF_DL_MinRate_GUs} and Fig. \ref{fig:RateCDF_DL_MinRate_UAVs}.  On the other hand,  the SCO-ICBA scheme when applied on COmMIMO ZF transmission can achieve the best UAVs' downlink rate performance at the price of the GUs performance.

Fig. \ref{fig:RateCDF_UL_MinRate} demonstrates the uplink rates performance comparison between SCO-IIA and SCO-ICBA schemes when the worst user's  uplink rate is optimized assuming both COmMIMO and CFmMIMO networks under the considered receive beamforming techniques for GUs in Fig. \ref{fig:RateCDF_UL_MinRate_GUs} and  UAVs in Fig. \ref{fig:RateCDF_UL_MinRate_UAVs}. Particularly, Fig. \ref{fig:RateCDF_UL_MinRate_GUs} illustrates that, in the case of optimizing the worst user's rate, the best GUs' uplink rate performance can be achieved when the SCO-IIA is employed with the CFmMIMO network under PZF reception. At the same time, SCO-IIA with PZF and CFmMIMO guarantees at relatively good UAVs' performance as depicted from Fig. \ref{fig:RateCDF_UL_MinRate_UAVs}. On the other hand, both SCO-IIA and SCO-ICBA schemes with COmMIMO under ZF reception improves the UAVs performance significantly at the price of the GUs performance as noticed from Fig. \ref{fig:RateCDF_UL_MinRate_UAVs} and Fig. \ref{fig:RateCDF_UL_MinRate_GUs}.

\begin{figure*}[!t]
 \centering
 \captionsetup{font=small}
\begin{subfigure}[b]{0.38\linewidth}
\centering
\captionsetup{font=small}
  \includegraphics[width=\columnwidth]{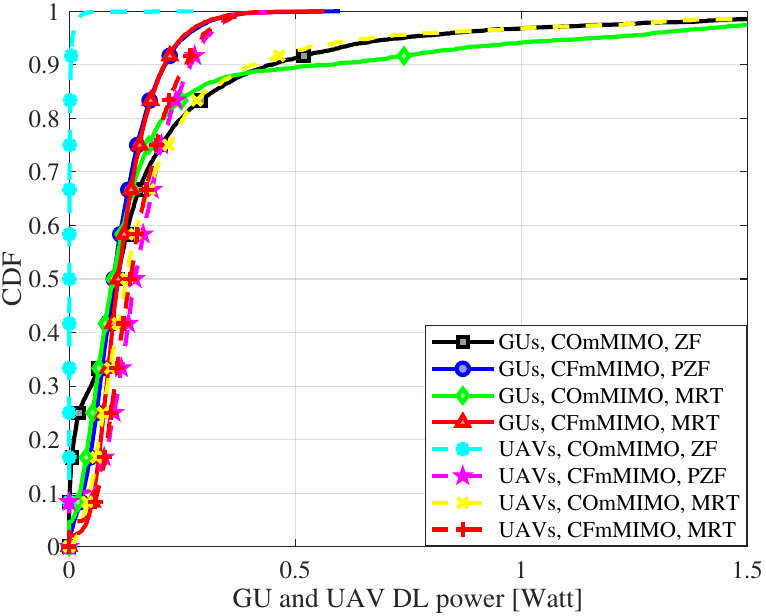} 
  \caption{SCO-IIA}
  \label{fig:PowerCDF_DL_MinRate_IIA}
\end{subfigure}\hfil 
\begin{subfigure}[b]{0.38\linewidth}
\centering
\captionsetup{font=small}
  \includegraphics[width=\columnwidth]{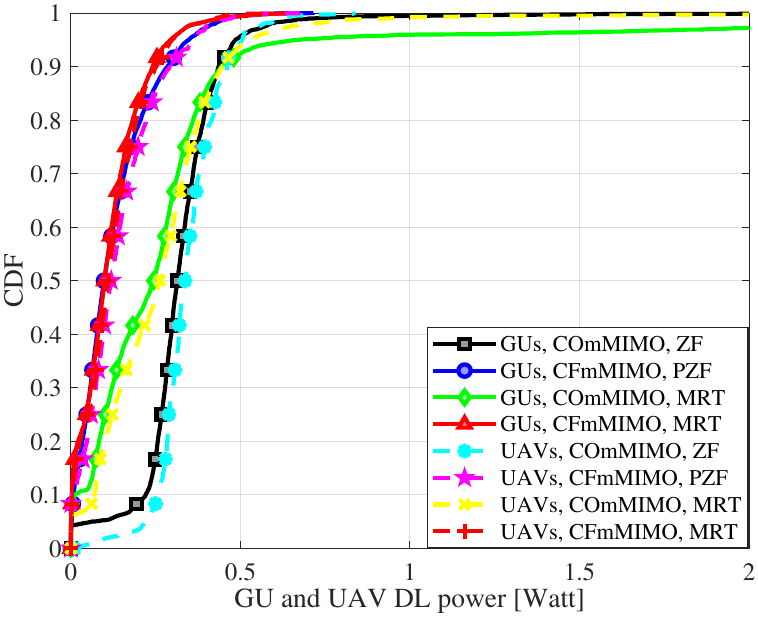}
  \caption{SCO-ICBA}
  \label{fig:PowerCDF_DL_MinRate_ICBA}
\end{subfigure} 
\caption{Downlink power consumption comparison between GUs and UAVs when both SCO-IIA and SCO-ICBA schemes are applied for minimum rate maximization under COmMIMO with ZF transmission, CFmMIMO with PZF transmission, COmMIMO with MRT, and CFmMIMO with MRT.} 
\label{fig:PowerCDF_DL_MinRate}
\end{figure*}

Fig. \ref{fig:PowerCDF_DL_MinRate} displays the downlink power consumption comparison between GUs and UAVs when the worst user's rate is optimized using both SCO-IIA and SCO-ICBA schemes. If we compare Fig. \ref{fig:PowerCDF_DL_MinRate} with Fig. \ref{fig:PowerCDF_DL_SumRate}, we can note that the gap between the CDF plot of the power consumed by GUs and that of UAVs when the objective of the considered optimization problem is to maximize the minimum user's rate (i.e., Fig. \ref{fig:PowerCDF_DL_MinRate}) is reduced in most cases when compared to the gap when the sum rate is the objective to be optimized (i.e., Fig. \ref{fig:PowerCDF_DL_SumRate}). Furthermore, it is noteworthy that maximizing the minimum rate has led to a reduction of power consumption of most UAVs and GUs in all the considered beamforming schemes with both SCO-IIA and SCO-ICBA for both CFmMIMO and COmMIMO.

\subsection{Comparative Summary and Discussion}
\begin{table*} [!t]
\centering
\captionsetup{font=small}
\caption{Performance comparison summary between SCO-IIA and SCO-ICBA when applied on downlink and uplink of COmMIMO with ZF, COmMIMO with MRT/MRC, CFmMIMO with PZF, and CFmMIMO with MRT/MRC for both sum rate and minimum rate optimization.}
\resizebox{1.4 \columnwidth}{!} {%
\begin{tabular}{|l|l|l|l|l|l|l|l|}
\hline
\multicolumn{1}{|c|}{scheme} & \multicolumn{1}{c|}{objective} & \multicolumn{1}{c|}{downlink/uplink} & \multicolumn{1}{c|}{network} & \multicolumn{1}{c|}{beamforming} & \multicolumn{1}{c|}{GUs $95\% R$ [Mbps]} & \multicolumn{1}{c|}{UAVs $95\% R$ [Mbps]} & \multicolumn{1}{c|}{$N_{it}$} \\ 
\hline
\multicolumn{1}{|c|}{} & \multicolumn{1}{c|}{} & \multicolumn{1}{c|}{} & \multicolumn{1}{c|}{COmMIMO} & \multicolumn{1}{c|}{ZF} & \multicolumn{1}{c|}{0} & \multicolumn{1}{c|}{38.11}  &  \multicolumn{1}{c|}{2} \\ 
\cline{5-8}
\multicolumn{1}{|c|}{} & \multicolumn{1}{c|}{} & \multicolumn{1}{c|}{} & \multicolumn{1}{c|}{} & \multicolumn{1}{c|}{MRT} & \multicolumn{1}{c|}{0} & \multicolumn{1}{c|}{11.45} &  \multicolumn{1}{c|}{5} \\ 
\cline{4-8}
\multicolumn{1}{|c|}{} & \multicolumn{1}{c|}{} & \multicolumn{1}{c|}{downlink} & \multicolumn{1}{c|}{CFmMIMO} & \multicolumn{1}{c|}{PZF} & \multicolumn{1}{c|}{4.15} & \multicolumn{1}{c|}{0} & \multicolumn{1}{c|}{2} \\ 
\cline{5-8}
\multicolumn{1}{|c|}{} & \multicolumn{1}{c|}{sum rate} & \multicolumn{1}{c|}{} & \multicolumn{1}{c|}{} & \multicolumn{1}{c|}{MRT} & \multicolumn{1}{c|}{3.27} & \multicolumn{1}{c|}{0} & \multicolumn{1}{c|}{2} \\ 
\cline{3-8}
\multicolumn{1}{|c|}{} & \multicolumn{1}{c|}{} & \multicolumn{1}{c|}{} & \multicolumn{1}{c|}{COmMIMO} & \multicolumn{1}{c|}{ZF} & \multicolumn{1}{c|}{0} & \multicolumn{1}{c|}{33.35}  &  \multicolumn{1}{c|}{2} \\ 
\cline{5-8}
\multicolumn{1}{|c|}{} & \multicolumn{1}{c|}{} & \multicolumn{1}{c|}{} & \multicolumn{1}{c|}{} & \multicolumn{1}{c|}{MRC} & \multicolumn{1}{c|}{0} & \multicolumn{1}{c|}{3.16}  & \multicolumn{1}{c|}{2} \\ 
\cline{4-8}
\multicolumn{1}{|c|}{} & \multicolumn{1}{c|}{} & \multicolumn{1}{c|}{uplink} & \multicolumn{1}{c|}{CFmMIMO} & \multicolumn{1}{c|}{PZF} & \multicolumn{1}{c|}{7.14} & \multicolumn{1}{c|}{10.24}  &  \multicolumn{1}{c|}{2} \\ 
\cline{5-8}
\multicolumn{1}{|c|}{SCO-IIA} & \multicolumn{1}{c|}{} & \multicolumn{1}{c|}{} & \multicolumn{1}{c|}{} & \multicolumn{1}{c|}{MRC} & \multicolumn{1}{c|}{0.61} & \multicolumn{1}{c|}{1.83} & \multicolumn{1}{c|}{2} \\ 
\cline{2-8}
\multicolumn{1}{|c|}{} & \multicolumn{1}{c|}{} & \multicolumn{1}{c|}{} & \multicolumn{1}{c|}{COmMIMO} & \multicolumn{1}{c|}{ZF} & \multicolumn{1}{c|}{0} & \multicolumn{1}{c|}{1.76} & \multicolumn{1}{c|}{5} \\ 
\cline{5-8}
\multicolumn{1}{|c|}{} & \multicolumn{1}{c|}{} & \multicolumn{1}{c|}{} & \multicolumn{1}{c|}{} & \multicolumn{1}{c|}{MRT} & \multicolumn{1}{c|}{2.72} & \multicolumn{1}{c|}{11.92}  & \multicolumn{1}{c|}{68} \\ 
\cline{4-8}
\multicolumn{1}{|c|}{} & \multicolumn{1}{c|}{} & \multicolumn{1}{c|}{downlink} & \multicolumn{1}{c|}{CFmMIMO} & \multicolumn{1}{c|}{PZF} & \multicolumn{1}{c|}{8.47} & \multicolumn{1}{c|}{0} &  \multicolumn{1}{c|}{17} \\ 
\cline{5-8}
\multicolumn{1}{|c|}{} & \multicolumn{1}{c|}{min rate} & \multicolumn{1}{c|}{} & \multicolumn{1}{c|}{} & \multicolumn{1}{c|}{MRT} & \multicolumn{1}{c|}{7.69} & \multicolumn{1}{c|}{4.58} &  \multicolumn{1}{c|}{52} \\ 
\cline{3-8}
\multicolumn{1}{|c|}{} & \multicolumn{1}{c|}{} & \multicolumn{1}{c|}{} & \multicolumn{1}{c|}{COmMIMO} & \multicolumn{1}{c|}{ZF} & \multicolumn{1}{c|}{0.05} & \multicolumn{1}{c|}{32.63}  & \multicolumn{1}{c|}{5} \\ 
\cline{5-8}
\multicolumn{1}{|c|}{} & \multicolumn{1}{c|}{} & \multicolumn{1}{c|}{} & \multicolumn{1}{c|}{} & \multicolumn{1}{c|}{MRC} & \multicolumn{1}{c|}{0} & \multicolumn{1}{c|}{11.01} & \multicolumn{1}{c|}{57} \\ 
\cline{4-8}
\multicolumn{1}{|c|}{} & \multicolumn{1}{c|}{} & \multicolumn{1}{c|}{uplink} & \multicolumn{1}{c|}{CFmMIMO} & \multicolumn{1}{c|}{PZF} & \multicolumn{1}{c|}{8.7} & \multicolumn{1}{c|}{11.41}  &  \multicolumn{1}{c|}{10} \\ 
\cline{5-8}
\multicolumn{1}{|c|}{} & \multicolumn{1}{c|}{} & \multicolumn{1}{c|}{} & \multicolumn{1}{c|}{} & \multicolumn{1}{c|}{MRC} & \multicolumn{1}{c|}{2.56} & \multicolumn{1}{c|}{5.58} &  \multicolumn{1}{c|}{36} \\ 
\hline
\multicolumn{1}{|c|}{} & \multicolumn{1}{c|}{} & \multicolumn{1}{c|}{} & \multicolumn{1}{c|}{COmMIMO} & \multicolumn{1}{c|}{ZF} & \multicolumn{1}{c|}{0} & \multicolumn{1}{c|}{38.66} &  \multicolumn{1}{c|}{2} \\ 
\cline{5-8}
\multicolumn{1}{|c|}{} & \multicolumn{1}{c|}{} & \multicolumn{1}{c|}{} & \multicolumn{1}{c|}{} & \multicolumn{1}{c|}{MRT} & \multicolumn{1}{c|}{0} & \multicolumn{1}{c|}{10.46} &  \multicolumn{1}{c|}{13} \\ 
\cline{4-8}
\multicolumn{1}{|c|}{} & \multicolumn{1}{c|}{} & \multicolumn{1}{c|}{downlink} & \multicolumn{1}{c|}{CFmMIMO} & \multicolumn{1}{c|}{PZF} & \multicolumn{1}{c|}{4.04} & \multicolumn{1}{c|}{0} &  \multicolumn{1}{c|}{16}\\ 
\cline{5-8}
\multicolumn{1}{|c|}{} & \multicolumn{1}{c|}{sum rate} & \multicolumn{1}{c|}{} & \multicolumn{1}{c|}{} & \multicolumn{1}{c|}{MRT} & \multicolumn{1}{c|}{0} & \multicolumn{1}{c|}{0} & \multicolumn{1}{c|}{25} \\ 
\cline{3-8}
\multicolumn{1}{|c|}{} & \multicolumn{1}{c|}{} & \multicolumn{1}{c|}{} & \multicolumn{1}{c|}{COmMIMO} & \multicolumn{1}{c|}{ZF} & \multicolumn{1}{c|}{0} & \multicolumn{1}{c|}{32.8} &  \multicolumn{1}{c|}{2} \\ 
\cline{5-8}
\multicolumn{1}{|c|}{} & \multicolumn{1}{c|}{} & \multicolumn{1}{c|}{} & \multicolumn{1}{c|}{} & \multicolumn{1}{c|}{MRC} & \multicolumn{1}{c|}{0} & \multicolumn{1}{c|}{0} &  \multicolumn{1}{c|}{18} \\ 
\cline{4-8}
\multicolumn{1}{|c|}{} & \multicolumn{1}{c|}{} & \multicolumn{1}{c|}{uplink} & \multicolumn{1}{c|}{CFmMIMO} & \multicolumn{1}{c|}{PZF} & \multicolumn{1}{c|}{9.31} & \multicolumn{1}{c|}{0} &  \multicolumn{1}{c|}{15} \\ 
\cline{5-8}
\multicolumn{1}{|c|}{SCO-ICBA} & \multicolumn{1}{c|}{} & \multicolumn{1}{c|}{} & \multicolumn{1}{c|}{} & \multicolumn{1}{c|}{MRC} & \multicolumn{1}{c|}{0} & \multicolumn{1}{c|}{0} & \multicolumn{1}{c|}{27} \\ 
\cline{2-8}
\multicolumn{1}{|c|}{} & \multicolumn{1}{c|}{} & \multicolumn{1}{c|}{} & \multicolumn{1}{c|}{COmMIMO} & \multicolumn{1}{c|}{ZF} & \multicolumn{1}{c|}{0} & \multicolumn{1}{c|}{36.25} &  \multicolumn{1}{c|}{7} \\ 
\cline{5-8}
\multicolumn{1}{|c|}{} & \multicolumn{1}{c|}{} & \multicolumn{1}{c|}{} & \multicolumn{1}{c|}{} & \multicolumn{1}{c|}{MRT} & \multicolumn{1}{c|}{0.95} & \multicolumn{1}{c|}{6.79} & \multicolumn{1}{c|}{94} \\ 
\cline{4-8}
\multicolumn{1}{|c|}{} & \multicolumn{1}{c|}{} & \multicolumn{1}{c|}{downlink} & \multicolumn{1}{c|}{CFmMIMO} & \multicolumn{1}{c|}{PZF} & \multicolumn{1}{c|}{7.47} & \multicolumn{1}{c|}{0.05} & \multicolumn{1}{c|}{85} \\ 
\cline{5-8}
\multicolumn{1}{|c|}{} & \multicolumn{1}{c|}{min rate} & \multicolumn{1}{c|}{} & \multicolumn{1}{c|}{} & \multicolumn{1}{c|}{MRT} & \multicolumn{1}{c|}{6.37} & \multicolumn{1}{c|}{3.07} & \multicolumn{1}{c|}{76} \\ 
\cline{3-8}
\multicolumn{1}{|c|}{} & \multicolumn{1}{c|}{} & \multicolumn{1}{c|}{} & \multicolumn{1}{c|}{COmMIMO} & \multicolumn{1}{c|}{ZF} & \multicolumn{1}{c|}{0} & \multicolumn{1}{c|}{32.23} &  \multicolumn{1}{c|}{6} \\ 
\cline{5-8}
\multicolumn{1}{|c|}{} & \multicolumn{1}{c|}{} & \multicolumn{1}{c|}{} & \multicolumn{1}{c|}{} & \multicolumn{1}{c|}{MRC} & \multicolumn{1}{c|}{0} & \multicolumn{1}{c|}{3.97} &  \multicolumn{1}{c|}{55} \\ 
\cline{4-8}
\multicolumn{1}{|c|}{} & \multicolumn{1}{c|}{} & \multicolumn{1}{c|}{uplink} & \multicolumn{1}{c|}{CFmMIMO} & \multicolumn{1}{c|}{PZF} & \multicolumn{1}{c|}{6.39} & \multicolumn{1}{c|}{7.49} &  \multicolumn{1}{c|}{25} \\ 
\cline{5-8}
\multicolumn{1}{|c|}{} & \multicolumn{1}{c|}{} & \multicolumn{1}{c|}{} & \multicolumn{1}{c|}{} & \multicolumn{1}{c|}{MRC} & \multicolumn{1}{c|}{3.87} & \multicolumn{1}{c|}{3.87} &  \multicolumn{1}{c|}{54} \\ 
\hline
\end{tabular}
}
\label{tab:discussion}
\end{table*}

\begin{figure*}[!t]
 \centering
 \captionsetup{font=small}
\begin{subfigure}[b]{0.38\linewidth}
\centering
\captionsetup{font=small}
  \includegraphics[width=\columnwidth]{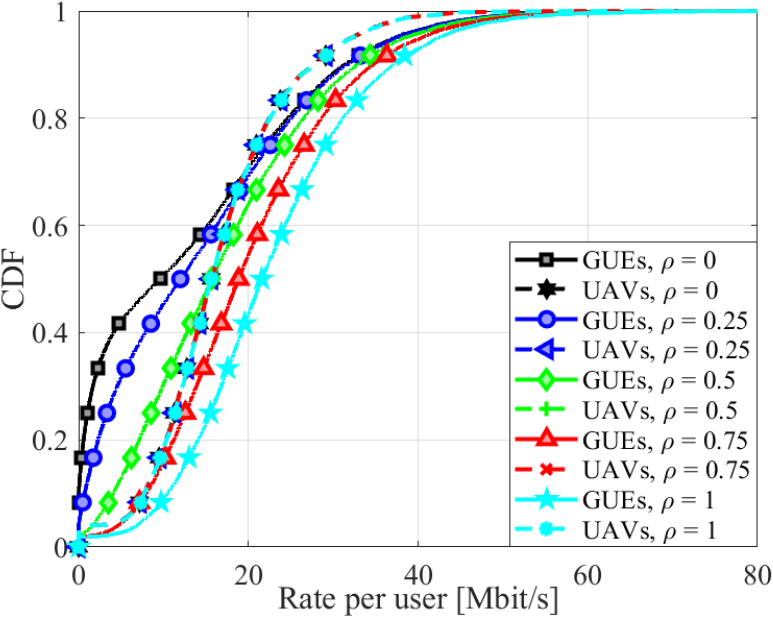} 
  \caption{MRT, downlink}
  \label{fig:RateCDF_DL_MRT_MinRate_IIA_rhoimpact}
\end{subfigure}\hfil 
\begin{subfigure}[b]{0.38\linewidth}
\centering
\captionsetup{font=small}
  \includegraphics[width=\columnwidth]{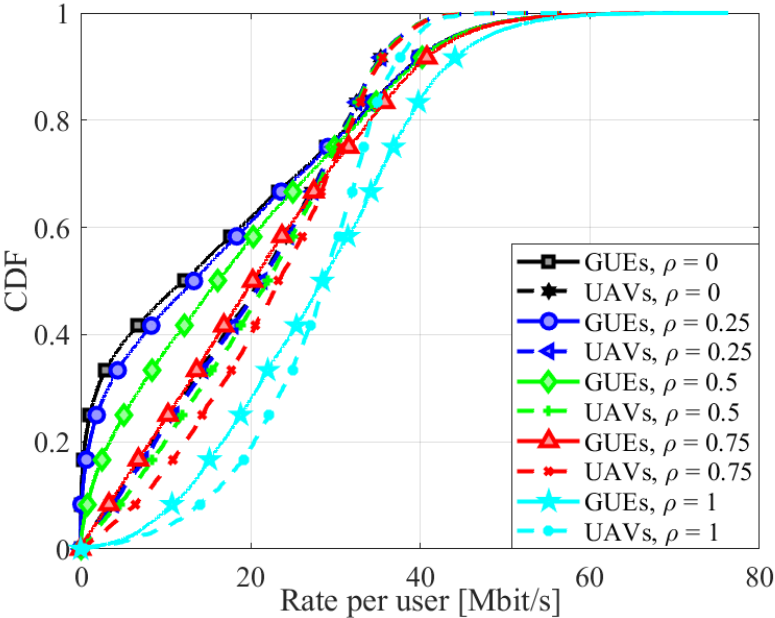}
  \caption{PZF, uplink}
  \label{fig:RateCDF_UL_PZF_MinRate_IIA_rhoimpact}
\end{subfigure}
\caption{Impact of small scale fading correlation on the performance of the proposed SCO-IIA for minimum rate maximization under CFmMIMO with MRT downlink and PZF uplink.} 
\label{fig:RateCDF_UL_PZF_MinRate_IIA}
\end{figure*}
Table \ref{tab:discussion} summarizes the obtained results by comparing the performance of the whole considered schemes in terms on the GUs 95\% likely rates (GUs 95\% R), the UAVs 95\% likely rates (UAVs 95\% R), and average number of iteration for convergence, $N_{it}$. In most cases, the COmMIMO can provide a relatively high values of UAVs 95\% R at the price of poor GUs 95\% R. On the other hand, the CFmMIMO improves the GUs 95\% R compared to the COmMIMO in most of the considered beamforming techniques. In the downlink, we recommend employing SCO-IIA scheme for minimum rate optimization in CFmMIMO with MRT as it converges after 52 iterations on average; and at the same time it provides GUs 95\% R of 7.69 Mbps and UAVs 95\% R of 4.07 Mbps. In the uplink, we recommend utilizing the SCO-IIA scheme for maximizing the minimum rate in CFmMIMO with PZF reception where it guarantees that the GUs 95\% R and the UAVs 95\% R are 8.425 Mbps and 11.32 Mbps, respectively; and converges after an average number of iterations of 10.

In this part, we discuss the impact of employing channel estimations synchronized with the large scale fading variation on the performance of the two recommended schemes. We assume that the samples of small scale fading coefficients $\boldsymbol{g}_{u,a}$ over consecutive coherence intervals, during which the large scale fading is not significantly changed, are correlated to each other by correlation factor $\rho$. Fig. \ref{fig:RateCDF_UL_PZF_MinRate_IIA} demonstrates the impact of $\rho$ on the performance of the recommended schemes for both downlink and uplink CFmMIMO. Specifically, we perform simulations for 250 different scenarios. In every scenario, we employ the recommended schemes and evaluate the performance of both UAVs and GUS in two coherence intervals at different values of $\rho$. It is clear that the performance of the UAVs is less sensitive to the value of $\rho$ compared to the GUs for both downlink as in Fig. \ref{fig:RateCDF_DL_MRT_MinRate_IIA_rhoimpact} and uplink as in Fig. \ref{fig:RateCDF_UL_PZF_MinRate_IIA_rhoimpact} because of the existence of the LOS path in the case of UAVs. Furthermore, it is noteworthy that the impact of $\rho$ on the performance of the GUs in both the downlink and the uplink is reduced by increasing the value of $\rho$, which is expected due to the less mobility of GUs. 

 \section{Conclusion}
This paper introduced the SCO-IIA and the SCO-ICBA schemes for power control in CFmMIMO network that support URLLC applications for both traditional GUs and UAVs. We consider optimizing the sum rate and the worst user's rate for both uplink and downlink. We benchmark the CFmMIMO against the COmMIMO while evaluating the performance of the proposed schemes. Our simulation results prove that the SCO-IIA scheme outperforms the SCO-ICBA scheme in terms of GUs rate, and UAVs rate. Furthemore, we found that the CFmMIMO outperforms COmMIMO specially for the GUs rate. Based on our experiments, we recommend the SCO-IIA scheme to optimize the minimum user's rate for CFmMIMO with MRT in the downlink, and PZF reception in the uplink.

\bibliographystyle{IEEEtran}
\bibliography{bibfile}

\begin{IEEEbiography}[{\includegraphics[width=1in,height=1.25in,clip,keepaspectratio]{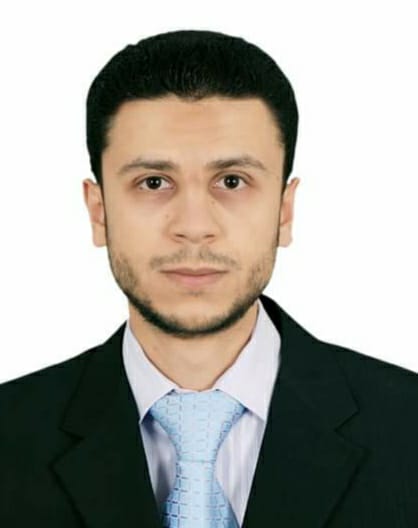}}]{Mohamed Elwekeil} received the B.Sc. degree in electronics and electrical communications engineering from the Faculty of Electronic Engineering, Menoufia University, Egypt, in 2007, and the M.Sc. and Ph.D. degrees in electronics and communications engineering from the Egypt-Japan University of Science and Technology (E-JUST), Alexandria, Egypt, in 2013 and 2016, respectively. He was with the Department of Electronics and Electrical Communications Engineering, Faculty of Electronic Engineering, Menoufia University since 2007, first as Teaching Assistant, and later, as Lecturer (Assistant Professor) and Associate Professor. From April 2014 to March 2015, he was a Research Intern at Alcatel-Lucent Bell N.V. (now Nokia), Antwerp, Belgium, where he was working on Wi-Fi optimization project. In October 2015, he joined Kyushu University, Fukuoka, Japan, as a Special Research Student, for a period of nine months. In April 2018, he joined the College of Information Engineering, Shenzhen University, Shenzhen, China, where he was working as a Postdoctoral Researcher, for a period of two year. In December 2020, he joined the Department of Electrical and Information Engineering, University of Cassino and Southern Lazio, Cassino, Italy, as a Postdoctoral Research Fellow, where his work was funded by the H2020 Marie Skłodowska-Curie Actions (MSCA) Individual Fellowships (IF) IUCCF. In June 2022, he has joined Nokia Bell Labs, Stuttgart, Germany as a visiting researcher as a part of the MSCA-IF fellowship. His research interests include radio resource management for wireless networks, spatial modulation, cell-free massive MIMO, machine learning applications in wireless networks, signal processing for communications, and earthquakes engineering. 

\end{IEEEbiography}

\begin{IEEEbiography}[{\includegraphics[width=1in,height=1.25in,clip,keepaspectratio]{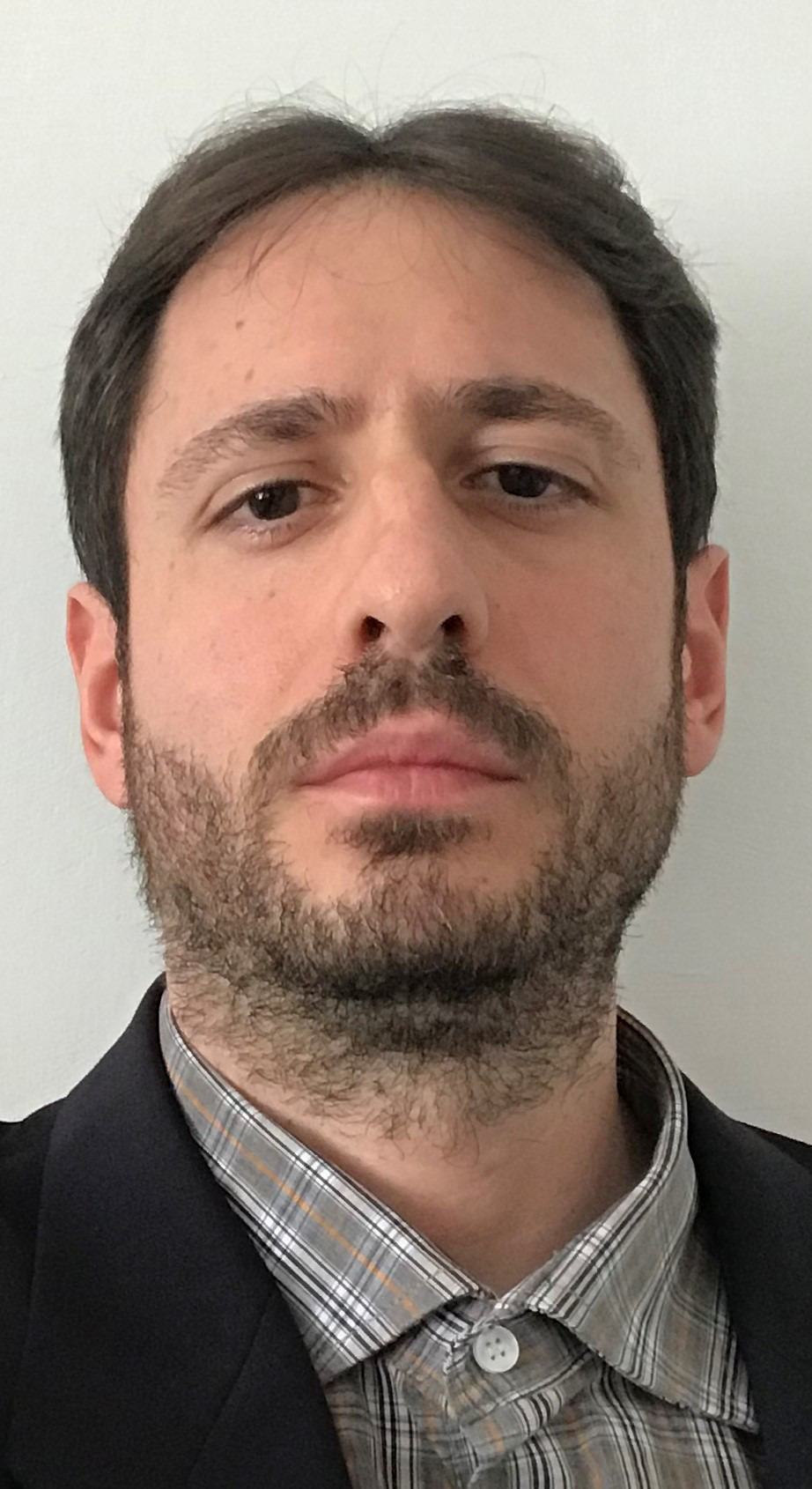}}]{Alessio Zappone} (SM'16) received his M.Sc. and Ph.D. both from the University of Cassino and Southern Lazio (Cassino, Italy). In 2012 he has worked with the Consorzio Nazionale Interuniversitario per le Telecomunicazioni (CNIT) in the framework of the FP7 EU-funded project TREND. From 2012 to 2016 he has been with the Dresden University of Technology, managing the project CEMRIN on energy-efficient resource allocation in wireless networks, funded by the German research foundation (DFG). In 2017 he was the recipient of the H2020 Marie Curie IF BESMART fellowship for experienced researchers, carried out at the LANEAS group of CentraleSupelec (Gif-sur-Yvette, France). Since 2019, he is a professor at the University of Cassino and Southern Lazio.	

His research interests lie in the area of communication theory and signal processing, with main focus on optimization techniques for resource allocation and energy efficiency maximization. For his research he received the Marconi Award of the IEEE Communications Society in 2021. Alessio serves as senior area editor for the \textsc{IEEE Signal Processing Letters}, as Editor of the \textsc{IEEE Transactions on Wireless Communications}, and has served as guest editor for the \textsc{IEEE Journal on Selected Areas on Communications}.

\end{IEEEbiography}

\begin{IEEEbiography}[{\includegraphics[width=1in,height=1.25in,clip,keepaspectratio]{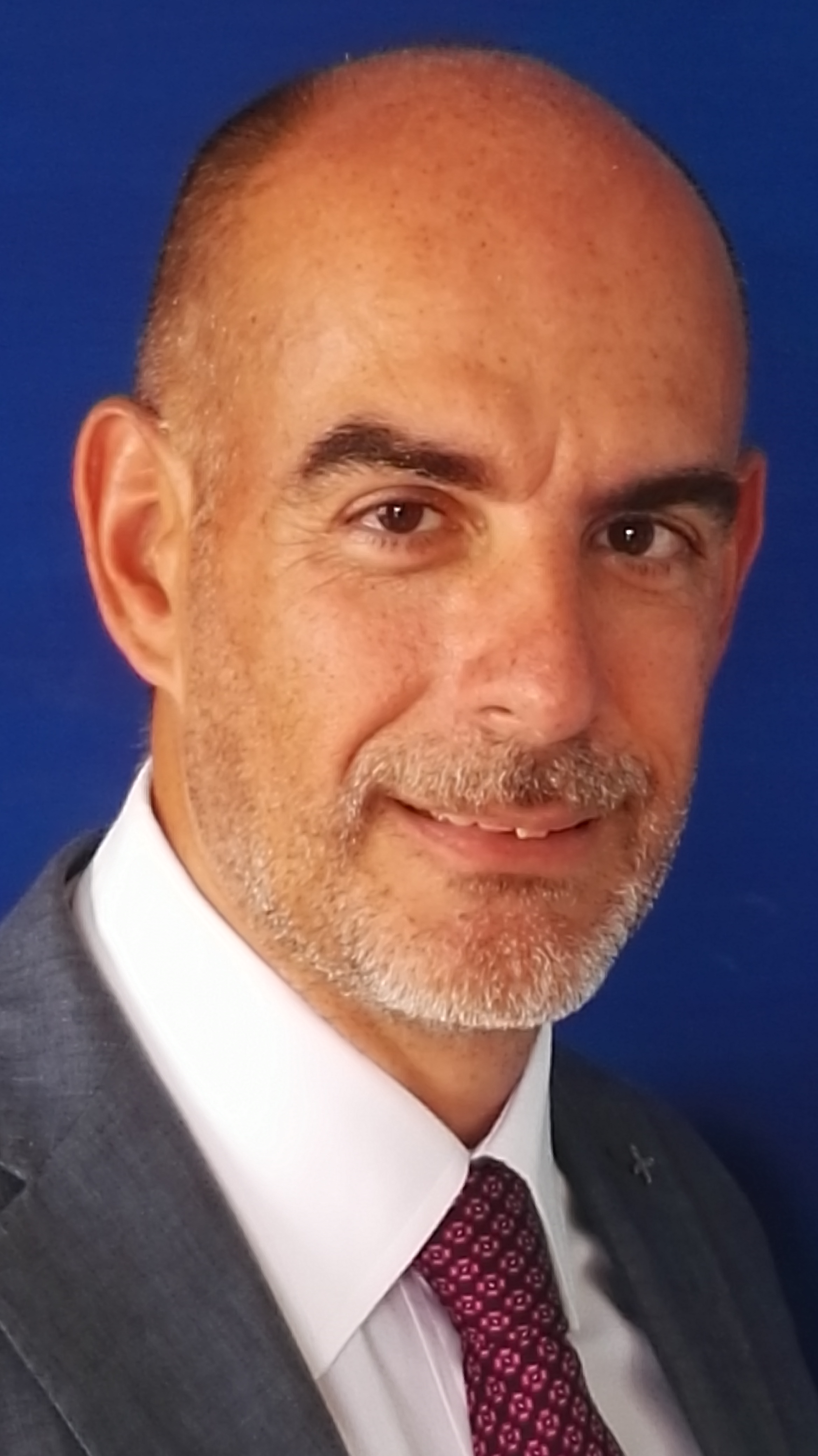}}]{Stefano Buzzi} (M'98-SM'07) joined the University of Cassino and Lazio Meridionale, Italy in 2000, first as an Assistant Professor, then as an Associate Professor (since 2002)  and, finally, since 2018, as a Full Professor. He received the M.Sc. degree (summa cum laude)in Electronic Engineering in 1994, and the Ph.D. degree in Electrical and Computer Engineering in 1999, both from the University of Naples “Federico II”.  He has had short-term research appointments at Princeton University, Princeton (NJ), USA in 1999, 2000, 2001 and 2006. He is a former Associate Editor of the IEEE Signal Processing Letters and of the IEEE Communications Letters, has been the guest editor of four IEEE JSAC special issues, and from 2014 to 2020 he has been an Editor for the IEEE Transactions on Wireless Communications. He also serves regularly as TPC member of several international conferences.  
Dr. Buzzi’s research interests are in the broad field of communications and signal processing, with emphasis on wireless communications and beyond-5G systems.  He is currently the Principal Investigator of the EU-funded Innovative Training Network project METAWIRELESS, on the application of metasurfaces to wireless communications. He has co-authored about 170 technical peer-reviewed journal and conference papers, and, among these, the highly cited paper "What will 5G be?", IEE JSAC, June 2014. 

\end{IEEEbiography}

\end{document}